\documentclass{revtex4}
\usepackage{amsmath,amsthm}
\usepackage{rotating}
\usepackage{blkarray}
\usepackage{booktabs}
\usepackage{mathtools}
\usepackage{multirow}
\usepackage{adjustbox}

\usepackage{graphicx}

\usepackage{amsfonts}
\newtheorem{theorem}{Theorem}[section]

\newtheorem{proposition}[theorem]{Proposition}

\theoremstyle{remark}
\newtheorem{remark}[theorem]{Remark}

\theoremstyle{definition}
\newtheorem{definition}[theorem]{Definition}

\theoremstyle{example}
\newtheorem{example}[theorem]{Example}

\theoremstyle{notation}

\newcommand{\bra}[1]{\langle#1|}
\newcommand{\ket}[1]{|#1\rangle}

\begin{document}

\title{Unitarily inequivalent local and global Fourier transforms in multipartite quantum systems}            
\author{C. Lei and A. Vourdas$^*$}
\affiliation{Department of Computer Science,\\
University of Bradford, \\
Bradford BD7 1DP, United Kingdom\\c.lei1@bradford.ac.uk\\a.vourdas@bradford.ac.uk}

\begin{abstract}
\textbf{\abstractname.} A multipartite system comprised of $n$ subsystems, each of which is described with `local variables' in ${\mathbb Z}(d)$ and with a $d$-dimensional Hilbert space $H(d)$, is considered.
Local Fourier transforms in each subsystem are defined and related phase space methods are discussed (displacement operators, Wigner and Weyl functions, etc).
A holistic view of the same system might be more appropriate in the case of strong interactions, which uses `global variables' in  ${\mathbb Z}(d^n)$ and a $d^n$-dimensional Hilbert space $H(d^n)$.
A global Fourier transform is then defined and related phase space methods are discussed.
The local formalism is compared and contrasted with the global formalism.
Depending on the values of $d,n$ the local Fourier transform is unitarily inequivalent or unitarily equivalent to the global Fourier transform.
Time evolution of the system in terms of both local and global variables, is discussed.
The formalism can be useful in the general area of Fast Fourier transforms.
\end{abstract}
\maketitle

\section{Introduction}
Entanglement and stronger than classical correlations in multipartite systems, are fundamental concepts in quantum mechanics (e.g., \cite{H}).
Even if the various components of the system are physically located far from each other,
strong correlations and strong interactions between them, weaken the concept of separate identity for each component .
 This motivates a comparison between the formalism of a multipartite system, with a holistic formalism of the same system that uses global quantities.
 
 We consider a finite quantum system with variables in ${\mathbb Z}(d)$ where $d$ is an odd integer, described by the $d$-dimensional Hilbert space $H(d)$ (e.g.\cite{V1,V11}).
 We also consider a multipartite system that consists of $n$ of these systems (which are possibly located far from each other).
In this system the positions and momenta take values in $[{\mathbb Z}(d)]^n={\mathbb Z}(d)\times ...\times{\mathbb Z}(d)$.
The system is described with the $d^n$-dimensional Hilbert space ${\mathfrak H}=H(d)\otimes ...\otimes H(d)$. 

In the case of strong correlations and strong interactions between the $n$ components we introduce a holistic approach and regard this as one system with variables in ${\mathbb Z}(d^n)$ and 
$d^n$-dimensional Hilbert space $H(d^n)$. We note that
\begin{itemize}

\item
The Hilbert space ${\mathfrak H}$ is isomorphic to the $H(d^n)$, because they both have the same dimension.
\item
There is a bijective map between the sets $[{\mathbb Z}(d)]^n$ and ${\mathbb Z}(d^n)$ given below in Eq.(\ref{16}) (in fact we can have many bijective maps between these two sets).
However the $[{\mathbb Z}(d)]^n$ as a ring is not isomorphic to the ring ${\mathbb Z}(d^n)$ (see Eq.(\ref{17}) below).

\end{itemize}
With this in mind, we study the following:
\begin{itemize}
\item
We define a local Fourier transform $F_L$ in the phase space $[{\mathbb Z}(d)]^n \times [{\mathbb Z}(d)]^n$ of the system when considered as $n$-component system.
We also define a global Fourier transform $F_G$ in the phase space ${\mathbb Z}(d^n) \times {\mathbb Z}(d^n)$ of the system when considered as a single system.
This has been introduced briefly in a different context in ref.\cite{V2}, and here it is studied as a problem in its own right and in connection with a global phase space formalism.
We show that depending on the values of $d,n$ the local Fourier transform is unitarily inequivalent (unitarily equivalent) to the global Fourier transform.
By that we mean that there exists no unitary transformation $U$ (there exists such a transformation $U$) so that $F_G=UF_LU^\dagger$.
This is discussed in section \ref{sec11} and in proposition \ref{pro1}.

\item
Starting from an orthonormal basis of `position states', we use local and global Fourier transforms to define local and global momentum states.
Some of the local momentum states are the same as the global momentum states as discussed in proposition \ref{pro17}.
We also define local position and momentum operators, and also global position and momentum operators.
We do numerical calculations of the time evolution for the case where the Hamiltonian is expressed in terms of local variables and also 
for the case where the Hamiltonian is expressed in terms of global variables (section \ref{GGG}).
For multipartite systems with strong interactions between the various components, it might be more appropriate to express the Hamiltonian in terms of the global variables.

\item
We define a local phase space formalism in $[{\mathbb Z}(d)]^n \times [{\mathbb Z}(d)]^n$ and a global phase space formalism in ${\mathbb Z}(d^n) \times {\mathbb Z}(d^n)$.
Displacements, Wigner and Weyl functions, etc, are defined in these two cases.
Density matrices which have only diagonal elements with respect to the position basis, have the same local and global Wigner function (proposition \ref{pro22}).
The difference between local and global Wigner functions, is contained entirely in the off-diagonal elements.

\item
Deviations of a density matrix $\rho$ from the corresponding factorisable density matrix ${\mathfrak R}(\rho)$ (defined in Eq.(\ref{1010})) are described 
with the matrices $R_L$, $\widetilde R_L$ and $R_G$, $\widetilde R_G$. They describe classical and quantum correlations  in the multipartite system described by $\rho$ (section \ref{sec35}). 

\item
Understanding of the relationship between global and local Fourier transforms and related phase space methods, might be useful in other areas like fast Fourier transforms. 
For $n=2$ we show that the global Fourier transform can be expressed in terms of many local Fourier transforms (section \ref{sec34}).
This is similar to the Cooley-Tukey formalism in fast Fourier transforms\cite{B1,B2,B3}.
The general area of Fast Fourier transforms (in a quantum or even classical context) is a potential application of the present formalism.

\item
In the case that the local and global Fourier transform are unitarily inequivalent (Eq.(\ref{60}) below), the concept of a multipartite system (and related concepts like entanglement) is fundamentally different from that of  a single quantum system.
But if they are unitarily equivalent (Eq.(\ref{59}) below), the distinction between a multipartite system and a single system is weak. Unitary equivalence means that with a change of basis one concept is transformed to another, and 
consequently there is no fundamental difference between the two. 
In this case, further work is needed  in order to clarify the correspondence between the two (especially of entanglement which is a concept applicable to a multipartite system but not to a single system).

\end{itemize}

In section 2 we review briefly the phase-space formalism for systems with finite Hilbert space\cite{V1,V11}.
In section 3 we apply this to each component of a $n$-partite system, and this is the `local formalism'.
In section 4 we define the global Fourier transform and discuss for which values of $d,n$ it is 
unitarily inequivalent to the local Fourier transform.
In section 5 we present the global phase space formalism and compare and contrast it with the local formalism.
In section 6, we present examples.
We conclude in section 7 with a discussion of our results.

\section{Background}

We  consider a quantum system (qudit) with variables in  the ring ${\mathbb Z}(d)$ of integers modulo $d$ where $d$ is an odd integer. 
$H(d)$ is the $d$-dimensional Hilbert space describing this system. 
There are well known technical differences between quantum systems with odd dimension $d$ and even dimension $d$ (e.g., \cite{EV0,EV1,EV2}).
In this paper we consider systems with odd dimension $d$.

Let $|X;j\rangle$ where $j\in {\mathbb Z}(d)$ be an orthonormal basis in $H(d)$.
The $X$ in the notation is not a variable, it simply indicates `position states'.
The finite Fourier transform $F$ is given by\cite{T}
\begin{eqnarray}\label{FF}
&&F=\frac{1}{\sqrt{d}}\sum _{j,k}\omega_d(jk) \ket{X;j}\bra{X;k};\;\;\;\omega_d(\alpha)=\exp \left (i\frac{2\pi\alpha}{d}\right);\;\;\;\alpha,j,k\in{\mathbb Z}(d)\nonumber\\
&&F^4={\bf 1};\;\;\;FF^{\dagger}={\bf 1}.
\end{eqnarray}
Its trace is\cite{V1}
\begin{eqnarray}\label{R1}
d=4m+1\rightarrow\;{\rm Tr}F=1;\nonumber\\
d=4m+3\rightarrow\;{\rm Tr}F=i.
\end{eqnarray}

We act with $F$ on position states and get the dual  basis
\begin{eqnarray}
&&\ket{P;j}=F\ket{X;j}.
\end{eqnarray}
The $P$ in the notation is not a variable, it simply indicates `momentum states'.

Using the relation
\begin{eqnarray}\label{100}
\frac{1}{d}\sum _{k}\omega_d[(j+\ell)k]=\delta(j,-\ell),
\end{eqnarray}
we show that $F^2$ is the parity operator around the origin:
\begin{eqnarray}\label{parity}
F^2=\frac{1}{d}\sum _{j,k,\ell}\omega_d[(j+\ell)k] \ket{X;j}\bra{X;\ell}=\sum _{j}\ket{X;j}\bra{X;-j}.
\end{eqnarray}

The phase space of this system is  ${\mathbb Z}(d)\times {\mathbb Z}(d)$ and in it we introduce the displacement operators
\begin{eqnarray}\label{A}
&&X^\beta = \sum _j\omega _d(-j\beta)|P; j \rangle \bra{P; j }=\sum _j\ket{X; j+\beta}\bra{X;j};\nonumber\\
&&Z^\alpha =\sum _j|P; j+\alpha \rangle \bra{P;j}= \sum _j\omega _d(\alpha j)|X; j\rangle\bra{X;j}=F X^\alpha F^\dagger;\nonumber\\
&&X^{d}=Z^{d}={\bf 1};\;\;\;X^\beta Z^\alpha= Z^\alpha X^\beta \omega_d(-\alpha \beta);\;\;\;\alpha, \beta\in {\mathbb Z}(d).
\end{eqnarray}
General displacement operators are the unitary operators
\begin{eqnarray}
&&D(\alpha, \beta)=Z^\alpha X^\beta \omega _d(-2^{-1}\alpha \beta);\;\;\;\;[D(\alpha, \beta)]^{\dagger}=D(-\alpha, -\beta);\nonumber\\
&&D(\alpha _1,\beta _1)D(\alpha _2,\beta _2)=D(\alpha _1+\alpha _2,\beta _1+\beta _2)\omega_d[2^{-1}(\alpha _1\beta _2 -\alpha _2\beta _1)].
\end{eqnarray}
The $2^{-1}=\frac{d+1}{2}$ is an integer in ${\mathbb Z}(d)$ with odd $d$, considered here.
The $D(\alpha, \beta)\omega (\gamma)$ form a representation of the 
Heisenberg-Weyl group.
We note that
\begin{eqnarray}
&&D(\alpha, \beta)\ket{X;  j}=\omega_{d}(2^{-1} \alpha  \beta +\alpha j)\ket{X; j+\beta};\nonumber\\
&&D(\alpha, \beta)\ket{P;  j}=\omega_{d}(-2^{-1} \alpha \beta - \beta j)\ket{P; j+\alpha}.
\end{eqnarray}

The 
\begin{eqnarray}\label{9}
{\cal X}=-i\frac{d}{2\pi}\log(Z);\;\;\;{\cal P}=i\frac{d}{2\pi}\log(X);\;\;\;F{\cal P}F^\dagger=-{\cal X}.
\end{eqnarray}
are $d\times d$ matrices which can be interpreted as position and momentum operators. 
The commutator $[{\cal X},{\cal P}]$ can be calculated (it is not $i{\bf 1}$) but it has no mathematical significance because the Heisenberg-Weyl group in this context is discrete, and the concept of generators is non-applicable.
Hamiltonians can be written as functions of these operators as $h({\cal X}, {\cal P})$.

\subsection{Wigner and Weyl functions}

The parity operator (around the point $(\gamma, \delta)$) is defined as
\begin{eqnarray}
{\mathfrak P}(\{\gamma, \delta\})=D(\gamma, \delta)F^2[D(\gamma, \delta)]^{\dagger};\;\;\;[{\mathfrak P}(\{\gamma, \delta\})]^2={\bf 1}.
\end{eqnarray}
It is related to the displacement operators through the Fourier transform 
\begin{eqnarray}\label{11}
&&{\mathfrak P}(\gamma,\delta)=\frac{1}{d}\sum_{\alpha, \beta}D(\alpha,\beta)\omega_d(\beta \gamma-\alpha \delta);\nonumber\\
&&D(\alpha,\beta)=\frac{1}{d}\sum_{\gamma, \delta}{\mathfrak P}(\gamma,\delta)\omega_d(-\beta \gamma+\alpha \delta).
\end{eqnarray}
If $\rho$ is a density matrix,  we define the Wigner function $W(\gamma, \delta)$ and the Weyl function $\widetilde W(\alpha, \beta)$ as:
\begin{eqnarray}
W(\gamma,\delta)={\rm Tr}[ \rho {\mathfrak P}(\gamma,\delta)];\;\;\;{\widetilde W}(\alpha,\beta)={\rm Tr}[ \rho D(\alpha,\beta)].
\end{eqnarray} 
From Eq.(\ref{11}) follows immediately that they are related to each other through the Fourier transform:
\begin{eqnarray}\label{120}
&&W(\gamma,\delta)=\frac{1}{d}\sum_{\alpha,\beta}{\widetilde W}(\alpha,\beta)\omega_d(\beta \gamma-\alpha \delta);\nonumber\\
&&{\widetilde W}(\alpha,\beta)=\frac{1}{d}\sum_{\gamma,\delta}W(\gamma,\delta)\omega_d(-\beta \gamma+\alpha \delta).
\end{eqnarray}
The following marginal properties of the Wigner function are well known for odd values of the dimension $d$ (e.g., \cite{V1}):
\begin{eqnarray}\label{A1}
&&\frac{1}{d}\sum _\gamma W(\gamma,\delta)=\bra{X;\delta}\rho\ket{X;\delta};\nonumber\\
&&\frac{1}{d}\sum _\delta W(\gamma,\delta)=\bra{P;\gamma}\rho\ket{P;\gamma};\nonumber\\
&&\frac{1}{d}\sum _{\gamma,\delta} W(\gamma,\delta)=1.
\end{eqnarray}

\section{Local phase space methods}
\subsection{Local Fourier transforms}

We consider a $n$-partite system comprised of $n$ components each of which is a qudit.
This system is described with the $d^n$-dimensional Hilbert space ${\mathfrak H}=H(d)\otimes ...\otimes H(d)$. 
Positions and momenta take values in $[{\mathbb Z}(d)]^n={\mathbb Z}(d)\times ...\times {\mathbb Z}(d)$.
If $\rho$ is the density matrix of the system, we use the notation
\begin{eqnarray}
\breve \rho_r={\rm Tr}_{i\ne r}\rho;\;\;\;r=0,...,n-1,
\end{eqnarray}
for the reduced density matrix describing the $r$-component of the system.
We also define the corresponding factorisable density matrix
\begin{eqnarray}\label{1010}
{\mathfrak R}(\rho)=\breve \rho_0\otimes...\otimes \breve \rho_{n-1};\;\;\;{\rm Tr}{\mathfrak R}(\rho)=1,
\end{eqnarray}
and the correlator
\begin{eqnarray}\label{101}
{\mathfrak C}(\rho)=\rho-{\mathfrak R}(\rho);\;\;\;{\rm Tr}{\mathfrak C}(\rho)=0.
\end{eqnarray}
For factorisable density matrices $\mathfrak R (\rho)=\rho$ and ${\mathfrak C}(\rho)=0$.
Below we compare quantities for $\rho$ with the corresponding quantities for $\mathfrak R (\rho)$.

We consider the basis
\begin{eqnarray}
\ket{X;j_0,...,j_{n-1}}=\ket{X;j_0}\otimes...\otimes\ket{X;j_{n-1}};\;\;\;j_r\in{\mathbb Z}(d).
\end{eqnarray}
called basis of position states.
We also consider the local Fourier transforms:
\begin{eqnarray}
F_L=F\otimes ...\otimes F;\;\;\;F_L^4={\bf 1};\;\;\;F_LF_L^{\dagger}={\bf 1}.
\end{eqnarray}
The index $L$ in the notation stands for local.
Acting with $F_L$ on the basis $\ket{X;j_0,...,j_{n-1}}$ we get the `local momentum states':
\begin{eqnarray}\label{103}
\ket{P_L;j_0,...,j_{n-1}}=F_L\ket{X;j_0,...,j_{n-1}}=\frac{1}{\sqrt{d^d}}\bigotimes _{r=0}^{n-1}\left [\sum _{k_r=0}^{d-1}\omega_{d}(j_rk_r)\ket{X;k_r}\right ]=\ket{P;j_0}\otimes...\otimes\ket{P;j_{n-1}}.
\end{eqnarray}

$F_L^2$ is a parity operator in the sense that
\begin{eqnarray}\label{103}
F_L^2\ket{X;j_0,...,j_{n-1}}=\ket{X;-j_0,...,-j_{n-1}}.
\end{eqnarray}

For later use we define the matrix elements of the correlator ${\mathfrak C}(\rho)$:
\begin{eqnarray}\label{BB1}
{\cal C}(X;j_0,...,j_{n-1})&=&\bra{X;j_0,...,j_{n-1}}{\mathfrak C}(\rho)\ket{X;j_0,...,j_{n-1}}\nonumber\\&=&\bra{X;j_0,...,j_{n-1}}\rho\ket{X;j_0,...,j_{n-1}}-\prod _{r=0}^{n-1}\bra{X;j_r}\breve \rho_r \ket{X;j_r},
\end{eqnarray}
and
\begin{eqnarray}\label{BB2}
{\cal C}(P_L;j_0,...,j_{n-1})&=&\bra{P_L;j_0,...,j_{n-1}}{\mathfrak C}(\rho)\ket{P_L;j_0,...,j_{n-1}}\nonumber\\&=&\bra{P_L;j_0,...,j_{n-1}}\rho\ket{P_L;j_0,...,j_{n-1}}-\prod _{r=0}^{n-1}\bra{P;j_r}\breve \rho_r \ket{P;j_r}.
\end{eqnarray}
Then 
\begin{eqnarray}
{\rm Tr}{\mathfrak C}(\rho)=\sum _{j_0,...,j_{n-1}}{\cal C}(X;j_0,...,j_{n-1})=\sum _{j_0,...,j_{n-1}}{\cal C}(X;j_0,...,j_{n-1})=0.
\end{eqnarray}
For factorisable density matrices ${\cal C}(X;j_0,...,j_{n-1})={\cal C}(P_L;j_0,...,j_{n-1})=0$.

\subsection{Displacements in $[{\mathbb Z}(d)\times {\mathbb Z}(d)]^n$}

The phase space of the system is  $[{\mathbb Z}(d)\times {\mathbb Z}(d)]^n$
and local displacement operators in it are defined as
\begin{eqnarray}\label{XX}
{\mathfrak X}_L(\{\beta_r\}) &=& \sum _{j_r}\omega _{d}(-\beta _0j_0-...-\beta_{n-1}j_{n-1})|P_L; j_0,...,j_{n-1}\rangle \bra{P_L; j_0,...,j_{n-1}}\nonumber\\&=&\sum _{j_r}\ket{X; j_0+\beta_0,...,j_{n-1}+\beta_{n-1}}\bra{X;j_0,...,j_{n-1}}\nonumber\\&=&
X^{\beta_0}\otimes...\otimes X^{\beta_{n-1}},
\end{eqnarray}
where $r=0,...,n-1$, and 
\begin{eqnarray}\label{A}
{\mathfrak Z}_L(\{\alpha_r\})& =&\sum _{j_r}|P_L; j_0+\alpha _0,...,j_{n-1}+\alpha_{n-1}\rangle \bra{P_L;j_0,...,j_{n-1}}\nonumber\\&=& 
\sum _{j_r}\omega _{d}(\alpha _0j_0+...+\alpha_{n-1}j_{n-1})|X; j_0,...,j_{n-1}\rangle\bra{X;j_0,...,j_{n-1}}\nonumber\\&=&
Z^{\alpha_0}\otimes...\otimes Z^{\alpha_{n-1}}=F_L{\mathfrak X}_L(\{\alpha_r\}) F_L ^\dagger.
\end{eqnarray}

Since ${\mathfrak Z}_L(\{\alpha_r\}){\mathfrak Z}_L(\{\gamma_r\})={\mathfrak Z}_L(\{\alpha_r+\gamma_r\})$, the ${\mathfrak Z}_L(\{\alpha_r\})$ form a representation of $[{\mathbb Z}(d)]^n$ as an additive group
The same is true for the ${\mathfrak X}_L(\{\beta_r\})$. Also
\begin{eqnarray}\label{26}
&&{\mathfrak X}_L(\{\beta_r\}){\mathfrak Z}_L(\{\alpha_r\})={\mathfrak Z}_L(\{\alpha_r\}){\mathfrak X}_L(\{\beta_r\})
\omega_d[-(\alpha_0\beta _0+...+\alpha_{n-1}\beta_{n-1})]\nonumber\\
&&[{\mathfrak X}_L(\{\beta_r\})]^d=[{\mathfrak Z}_L(\{\alpha_r\})]^d={\bf 1}.
\end{eqnarray}
Using the notation
\begin{eqnarray}
\{\alpha_r,\beta_r\}=\{a_0, ...,a_{n-1},\beta_0,...,\beta_{n-1}\},
\end{eqnarray}
general local displacement operators are defined as
\begin{eqnarray}\label{28}
D_L(\{\alpha_r,\beta_r\})&=&{\mathfrak Z}_L(\{\alpha_r\}){\mathfrak X}_L(\{\beta_r\})\omega_d[-2^{-1}(\alpha_0\beta _0+...+\alpha_{n-1}\beta_{n-1})]
\nonumber\\&=& D(\alpha_0, \beta_0)\otimes...\otimes D (\alpha_{n-1}, \beta_{n-1})\;\;\;\alpha_r,\beta_r\in{\mathbb Z}(d).
\end{eqnarray}
The $D_L(\{\alpha_r,\beta_r\})\omega (\{\gamma_r\})$ form a representation of the Heisenberg-Weyl group of displacements in the phase space $[{\mathbb Z}(d)\times {\mathbb Z}(d)]^n$.

The local parity operator (around the point $\{\gamma_r,\delta_r\}$ in the phase space  $[{\mathbb Z}(d)\times {\mathbb Z}(d)]^n$) is defined as
\begin{eqnarray}
{\mathfrak P}_L(\{\gamma_r,\delta_r\})&=&D_L(\{\gamma_r,\delta_r\})F_L^2[D_L(\{\gamma_r,\delta_r\})]^{\dagger}={\mathfrak P}(\gamma_0, \delta_0)\otimes...\otimes {\mathfrak P} (\gamma_{n-1}, \delta_{n-1})\nonumber\\
\left [{\mathfrak P}_L(\{\gamma_r,\delta_r\})\right ]^2&=&{\bf 1}.
\end{eqnarray}
It is  related to the local displacement operators through the Fourier transform
\begin{eqnarray}\label{12}
&&{\mathfrak P}_L(\{\gamma_r,\delta_r\})=\frac{1}{d^n}\sum_{\{\alpha_r,\beta_r\}}D_L(\{\alpha_r,\beta_r\})\omega_d\left [\sum_{r=0}^{n-1}(\beta _r\gamma_r-\alpha _r\delta_r)\right ];\nonumber\\
&&D_L(\{\alpha_r,\beta_r\})=\frac{1}{d^n}\sum_{\{\gamma_r,\delta_r\}}{\mathfrak P}_L(\{\gamma_r,\delta_r\})\omega_d\left [\sum_{r=0}^{n-1}(-\beta _r\gamma_r+\alpha _r\delta_r)\right ].
\end{eqnarray}
The proof of this follows easily from Eq.(\ref{11}).

\subsection{Local Wigner and local Weyl functions in $[{\mathbb Z}(d)\times {\mathbb Z}(d)]^n$}

If $\rho$ is a density matrix,  we define the local Wigner function $W_L(\{\gamma_r, \delta_r\}|\rho)$ and the local Weyl function $\widetilde W_L(\{\alpha_r, \beta_r\}|\rho)$ as:
\begin{eqnarray}\label{WL}
W_L(\{\gamma_r,\delta_r\}|\rho)={\rm Tr}[ \rho {\mathfrak P}_L(\{\gamma_r,\delta_r\})];\;\;\;{\widetilde W_L}(\{\alpha_r,\beta_r\}|\rho)={\rm Tr}[ \rho D_L(\{\alpha_r,\beta_r\})].
\end{eqnarray} 
From Eq.(\ref{12}) follows immediately that they are related to each other through the Fourier transform:
\begin{eqnarray}
&&W_L(\{\gamma_r,\delta_r\}|\rho)=\frac{1}{d^n}\sum_{\{\alpha_r,\beta_r\}}{\widetilde W_L}(\{\alpha_r,\beta_r\}|\rho)\omega_d\left [\sum_{r=0}^{n-1}(\beta _r\gamma_r-\alpha _r\delta_r)\right ];\nonumber\\
&&{\widetilde W_L}(\{\alpha_r,\beta_r\}|\rho)=\frac{1}{d^n}\sum_{\{\gamma_r,\delta_r\}}W_L(\{\gamma_r,\delta_r\}|\rho)\omega_d\left [\sum_{r=0}^{n-1}(-\beta _r\gamma_r+\alpha _r\delta_r)\right ].
\end{eqnarray}

\section{Global Fourier tarnsforms}
\subsection{A bijective map between the non-isomorphic rings $[{\mathbb Z}(d)]^n$ and ${\mathbb Z}(d^n)$}

We consider a bijective map between $[{\mathbb Z}(d)]^n$ and ${\mathbb Z}(d^n)$ as follows.
We first take each $j_r\in {\mathbb Z}(d)$ and $\widehat j\in {\mathbb Z}(d^n)$ in the `periods' 
\begin{eqnarray}\label{47}
\left [-\frac{d-1}{2}, \frac{d-1}{2}\right];\;\;\;
\left [-\frac{d^n-1}{2}, \frac{d^n-1}{2}\right],
\end{eqnarray}
correspondingly (for odd $d$).
We introduce the bijective map
\begin{eqnarray}\label{16}
j= (j_0,...,j_{d-1})\;\leftrightarrow\;\widehat j=j_0+j_1d+...+j_{n-1}d^{n-1}.
\end{eqnarray}
We then take each $j_r$ modulo $d$ and the $\widehat j$ modulo $d^n$, and we get a bijective map from $[{\mathbb Z}(d)]^n$ to ${\mathbb Z}(d^n)$.
Numbers in ${\mathbb Z}(d^n)$ will be denoted with a `hat', so that it is clear whether a number belongs to ${\mathbb Z}(d)$ or to ${\mathbb Z}(d^n)$.

The Hilbert space ${\mathfrak H}$ is isomorphic to $H(d^n)$ (a $d^n$-dimensional Hilbert space describing systems with variables in ${\mathbb Z}(d^n)$). 
But the $[{\mathbb Z}(d)]^n$ as a ring (with addition and multiplication componentwise), is not isomorphic to the ring ${\mathbb Z}(d^n)$ because addition and multiplication is different, and consequently our `local formalism' is different from our `global formalism'. Indeed
\begin{eqnarray}\label{17}
\widehat j+\widehat k\ne \widehat {j+k};\;\;\;\widehat j\widehat k\ne \widehat {jk}.
\end{eqnarray}
The sum is different because $\widehat j+\widehat k$ in ${\mathbb Z}(d^n)$ has  the `carry' rule and the $r$-component might be $j_r+k_r+1$ rather than $j_r+k_r$ .
In contrast, there is no `carry' rule in $[{\mathbb Z}(d)]^n$:
\begin{eqnarray}
j+k= (j_0+k_0,...,j_{d-1}+k_{d-1})\;\leftrightarrow\;\widehat {j+k}=(j_0+k_0)+(j_1+k_1)d+...+(j_{n-1}+k_{n-1})d^{n-1}.
\end{eqnarray}

Also multiplication in ${\mathbb Z}(d^n)$ is
\begin{eqnarray}\label{56}
\widehat j\widehat k=j_0k_0+d(j_1k_0+k_1j_0)+...+d^{n-1}(j_0k_{n-1}+...+j_{n-1}k_0).
\end{eqnarray}
The corresponding multiplication in $[{\mathbb Z}(d)]^n$ is
\begin{eqnarray}
(j_0,...,j_{n-1})\cdot (k_0,...,k_{n-1})=(j_0k_0,...,j_{n-1}k_{n-1}),
\end{eqnarray}
and with the bijective map in Eq.(\ref{16}) this corresponds to 
\begin{eqnarray}
\widehat {jk}=j_0k_0+d(j_1k_1)+...+d^{n-1}(j_{n-1}k_{n-1}).
\end{eqnarray}
It is seen that in general $\widehat j\widehat k\ne \widehat {jk}$ (but $\widehat 1\widehat k= \widehat {k}$).

\begin{example}
We consider the elements of ${\mathbb Z}(3)$ in the `period' $[-1,1]$ and the elements of ${\mathbb Z}(9)$ in the `period' $[-4,4]$.
A bijective map between $[{\mathbb Z}(3)]^2$ and ${\mathbb Z}(9)$ is as follows
\begin{eqnarray}
&&\widehat{(-1,-1)}=\widehat {-4};\;\;\widehat{(0,-1)}=\widehat {-3};\;\;\widehat{(1,-1)}= \widehat {-2};\;\;\widehat{(-1,0)}=\widehat {-1};\;\;\widehat{(0,0)}=\widehat 0;\nonumber\\
&&\widehat{(1,0)}= \widehat 1;\;\;\widehat{(-1,1)}=\widehat 2;\;\;\widehat{(0,1)}=\widehat 3;\;\;\widehat{(1,1)}= \widehat 4.
\end{eqnarray}
An example of addition that confirms Eq.(\ref{17}) is the following:
\begin{eqnarray}
&&\widehat {(1,1)}+\widehat {(1,1)}=\widehat 4+\widehat 4=\widehat {-1};\nonumber\\
&&\widehat{(1,1)+(1,1)}=\widehat{(-1,-1)}=\widehat {-4}.
\end{eqnarray}
An example of multiplication 
that confirms Eq.(\ref{17}) is the following:
\begin{eqnarray}
&&\widehat {(-1,0)}\cdot \widehat {(0,1)}=\widehat {-1}\cdot \widehat 3=\widehat {-3};\nonumber\\
&&\widehat{(-1,0)\cdot(0,1)}=\widehat{(0,0)}=\widehat 0 .
\end{eqnarray}

\end{example}

\begin{remark}\label{BB}
If $d_1,...,d_n$ are coprime to each other, then the ${\mathbb Z}(d_1)\times ...\times {\mathbb Z}(d_n)$ is isomorphic to ${\mathbb Z}(d_1\times...\times d_n)$.
We can define a bijective map
\begin{eqnarray}
(j_0,...,j_{n-1})\;\leftrightarrow\;j;\;\;\;j_r\in {\mathbb Z}(d_r);\;\;\;j\in {\mathbb Z}(d_1\times...\times d_n),
\end{eqnarray}
such that
\begin{eqnarray}
&&(j_0+k_0,...,j_{n-1}+k_{n-1})\;\leftrightarrow\;j+k;\nonumber\\
&&(j_0k_0,...,j_{n-1}k_{n-1})\;\leftrightarrow\;jk.
\end{eqnarray}
This is based on the Chinese remainder theorem, and has been used by Good \cite{G} in the context of fast Fourier transforms (see also \cite{B1,B2,B3}).
In a quantum context it has been use in \cite{V3,V1} for factorisation of a quantum system into subsystems. Here we consider the case $d_1=...=d_n$ and then the bijective map
of Eq.(\ref{16}) does not establish an isomorphism between the ring $[{\mathbb Z}(d)]^n$ and the ring  ${\mathbb Z}(d^n)$ (because of Eq.(\ref{17})).

\end{remark}

\subsection{Dual notation}

We use the following  dual notation for position states, based on the bijective map in Eq.(\ref{16}):
\begin{eqnarray}
\ket{X; j_0,...,j_{n-1}}=\ket{X;\widehat j}.
\end{eqnarray}
When local operators act on them we use addition and multiplication in $[{\mathbb Z}(d)]^n$, in connection with the phase space $[{\mathbb Z}(d)\times {\mathbb Z}(d)]^n$.
When global operators (defined below) act on them we use addition and multiplication  in ${\mathbb Z}(d^n)$, in connection with the phase space ${\mathbb Z}(d^n)\times {\mathbb Z}(d^n)$.

Analogous dual notation is used for all quantities. For example, the displacement operators  in Eq.(\ref{28}) can be denoted as
\begin{eqnarray}
D_L(\{\alpha_r,\beta_r\})=D_L(\widehat \alpha, \widehat \beta).
\end{eqnarray}
In some equations both notations appear together.

\subsection{Global Fourier transforms}\label{GF}
The global Fourier transform in ${\mathfrak H}$ is defined as:
\begin{eqnarray}
F_G&=&\frac{1}{\sqrt{d^n}}\sum _{{\widehat j},{\widehat k}}\omega_{d^n}(\widehat j\widehat k)\ket{X;j_0,...,j_{n-1}}\bra{X;k_0,...,k_{n-1}}.
\end{eqnarray}
The index $G$ in the notation stands for global. 
It is easily seen that
\begin{eqnarray}
F_G^4={\bf 1};\;\;\;F_GF_G^{\dagger}={\bf 1};\;\;\;F_G\ne F_L.
\end{eqnarray}

Acting with $F_G$ on the basis $\ket{X;j_0,...,j_{n-1}}$ (which we also denote as $\ket{X;\widehat j}$) we get the `global momentum states':
\begin{eqnarray}
&&\ket{P_G;\widehat j}=\ket{P_G;j_0,...,j_{n-1}}=F_G\ket{X;\widehat j}=\frac{1}{\sqrt{d^n}}\bigotimes _{r=0}^{n-1}\left [\sum _{k_r=0}^{d-1}\omega_{d^n}[(j_0d^r+..+j_{n-r-1}d^{n-1})k_r]\ket{X;k_r}\right ].
\end{eqnarray}
In the states $\ket{P_G;j_0,...,j_{n-1}}$, the coefficients $\omega_{d^n}[(j_0d^r+..+j_{n-r-1}d^{n-1})k_r]$ in the $r$-component depend on all $j_0,...,j_{n-1}$, and the term `global' refers to this.
Information from all components is needed, in order to determine these coefficients.
In the local Fourier transform of Eq.(\ref{103}), the coefficients $\omega_{d}(j_rk_r)$ in the $r$-component depend only on $j_r$. We note that
\begin{eqnarray}\label{44}
\langle P_L;\ell_0,...,\ell_{n-1}\ket{P_G;j_0,...,j_{n-1}}&=&\bra{ X;\ell_0,...,\ell_{n-1}}F_L^\dagger F_G\ket{X;j_0,...,j_{n-1}}\nonumber\\&=&\frac{1}{d^n}\sum_{\widehat k}\omega_{d^n}(\widehat j \widehat k)\omega_d[-(\ell_0k_0+...+\ell_{d-1}k_{n-1})].
\end{eqnarray}
and that
\begin{eqnarray}
|\langle X;\ell_0,...,\ell_{n-1}\ket{P_L;j_0,...,j_{n-1}}|^2=|\langle X;\ell_0,...,\ell_{n-1}\ket{P_G;j_0,...,j_{n-1}}|^2=\frac{1}{d^n}.
\end{eqnarray}

\begin{proposition}\label{pro17}
We take the elements of 
${\mathbb Z}(d)$ and the elements of 
${\mathbb Z}(d^n)$ in the `periods' of Eq.(\ref{47}).
Then
\begin{itemize}
\item[(1)]
The parity operator around the origin is the same in both the local and global formalism:
\begin{eqnarray}
F_G^2=F_L^2=\begin{pmatrix}
0&\cdots&0&1\\
0&\cdots&1&0\\
\vdots & \vdots &\vdots&\vdots\\
1&\cdots&0&0
\end{pmatrix}.
\end{eqnarray}
Here the matrix is in the position basis.

\item[(2)]
For any $n$
\begin{eqnarray}\label{456}
\ket{P_G;\widehat{-d^{n-1}}}=\ket{ P_L;-1, 0,...,0};\;\;\;\ket{P_G;0}=\ket{ P_L;0, 0,...,0};\;\;\;\ket{P_G;\widehat {d^{n-1}}}=\ket{ P_L;1, 0,...,0}.
\end{eqnarray} 
\item[(3)]
For $n=2$ we have the stronger result
\begin{eqnarray}\label{456}
\ket{P_G;\widehat{d\lambda}}=\ket{ P_L;\lambda,0};\;\;\;\lambda=-\frac{d-1}{2},...,\frac{d-1}{2}.
\end{eqnarray} 
At least $d$ of the global momentum states are equal to $d$ of the local momentum states.

\end{itemize}
\end{proposition}
\begin{proof}
\begin{itemize}
\item[(1)]
For ${\mathbb Z}_{d^n}$ Eq.(\ref{100}) becomes
\begin{eqnarray}
\frac{1}{{d^n}}\sum _{ \widehat k}\omega_{d^n}[(\widehat j+\widehat \ell )\widehat k] =\delta (\widehat j+\widehat \ell ,0);\;\;\;\widehat j, \widehat \ell\in {\mathbb Z}_{d^n}.
\end{eqnarray}
The ${\widehat j}+{\widehat \ell}=0$ implies $j_r+\ell _r=0$,
and we prove that
\begin{eqnarray}
F_G^2=\frac{1}{{d^d}}\sum _{\widehat j, \widehat k, \widehat \ell}\omega_{d^n}[(\widehat j+\widehat \ell )\widehat k] \ket{j_0,...,j_{n-1}}\bra{\ell_0,...,\ell_{n-1}}=
\sum_{j_0,...,j_{n-1}}\ket{j_0,...,j_{n-1}}\bra{-j_0,...,-j_{n-1}}=F_L^2.
\end{eqnarray}

\item[(2)]
Using Eq.(\ref{44}) we get
\begin{eqnarray}
\langle { P_L;1,...,0}\ket{P_G;\widehat{d^{n-1}}}&=&
\frac{1}{d^n}\sum_{\widehat k}\omega_{d^n}(\widehat{d^{n-1}}\widehat k)\omega_d(-k_0)\nonumber\\&=&\frac{1}{d^n}\sum_{\widehat k}\omega_{d^n}(d^{n-1}k_0)\omega_d(-k_0)=\frac{1}{d^n}\sum_{\widehat k}1=1.
\end{eqnarray}
In a similar way we prove that
\begin{eqnarray}
\ket{P_G;\widehat{-d^{n-1}}}=\ket{ P_L; -1, 0,...,0}.
\end{eqnarray} 
\item[(3)]
Using Eq.(\ref{44}) with $n=2$ and $\widehat k=k_0+dk_1$ we get
\begin{eqnarray}
\langle P_L;\lambda, 0\ket{P_G;\widehat{d\lambda}}&=&\frac{1}{d^2}\sum_{\widehat k}\omega_{d^2}(\widehat{d\lambda} \widehat k)\omega_d(-\lambda k_0)
=\frac{1}{d^2}\sum_{\widehat k}\omega_{d^2}(\widehat{d\lambda} \widehat k-d\lambda k_0)
\nonumber\\&=&
\frac{1}{d^2}\sum_{\widehat k}\omega_{d^2}(d^2\lambda k_1)=1.
\end{eqnarray}
$d\lambda$ takes values between $-\frac{d^2-1}{2}$ and $\frac{d^2-1}{2}$ and consequently $\lambda $ takes the values in Eq.(\ref{456}).

\end{itemize}
\end{proof}
For later use we define the matrix elements of the correlator ${\mathfrak C}(\rho)$:
\begin{eqnarray}\label{C1}
{\cal C}(X;\widehat j)=\bra{X;\widehat j}{\mathfrak C}(\rho)\ket{X;\widehat j};\;\;\;{\cal E}(P_G;\widehat j)=\bra{P_G;\widehat j}{\mathfrak C}(\rho)\ket{P_G;\widehat j}.
\end{eqnarray}
In both the `local formalism' and the `global formalism' the position states are the same and the momentum states are different.
Consequently the ${\cal E}(P_G;\widehat j)$ is different from the corresponding ${\cal C}(P_L;j_0,...,j_{n-1})$.

Then 
\begin{eqnarray}
{\rm Tr}{\mathfrak C}(\rho)=\sum _{\widehat j}{\cal C}(X;\widehat j)=\sum _{\widehat j}{\cal E}(P_G;\widehat j)=0.
\end{eqnarray}
For factorisable density matrices ${\cal C}(X;\widehat j)={\cal E}(P_G;\widehat j)=0$.

\subsection{Unitarily inequivalent local and global Fourier transforms}\label{sec11}

In this paper we use the following definition of unitary equivalence. Two square matrices $A,B$ are called unitarily equivalent if there exists a unitary matrix $U$ such that $A=UBU^\dagger$. Unitary equivalence is an equivalence relation, i.e., 
matrices which are unitarily inequivalent belong to different equivalence classes.
It is known\cite{E1,E2} that two normal $d\times d$ matrices $A,B$ are unitarily equivalent if and only if
\begin{eqnarray}\label{678}
||A||=||B||;\;\;\;{\rm Tr}(A^\eta)={\rm Tr}(B^\eta);\;\;\;\eta=1,...,d;\;\;\;||A||=\sqrt{\sum_{i,j}|A_{ij}|^2}.
\end{eqnarray}
We note that Specht's general theorem for unitary equivalence (e.g., \cite{E2}) reduces easily to the above criteria for the Fourier matrices which are unitary.

Some authors call the above unitary similarity, and they use the term unitary equivalence for the case where there exist two unitary matrices $U,V$ such that $A=UBV^\dagger$.

\begin{proposition}\label{pro1}
In an $n$-partite system that has Hilbert space with dimension $d^n$,
the $d^n\times d^n$ matrices $F_G$, $F_L$ are unitarily equivalent in the cases
\begin{eqnarray}\label{59}
&&d=4m+1;\nonumber\\
&&d=4m+3\;\;{\rm and}\;\;n=4N;\nonumber\\
&&d=4m+3\;\;{\rm and}\;\;n=4N+1.
\end{eqnarray}
The matrices $F_G$, $F_L$ are unitarily inequivalent in the cases
\begin{eqnarray}\label{60}
&&d=4m+3\;\;{\rm and}\;\;n=4N+2;\nonumber\\
&&d=4m+3\;\;{\rm and}\;\;n=4N+3.
\end{eqnarray}

\end{proposition}
\begin{proof}
The matrices $F_G$ and $F_L$ are unitary and therefore normal, and we use the criterion in Eq.(\ref{678}).
We first note that 
\begin{eqnarray}
||F_G||=||F_L||=d^n.
\end{eqnarray}
We next compare ${\rm Tr}(F_G^\eta )$ with ${\rm Tr}(F_L^\eta)$ for $\eta=1,...,d^n$. But
\begin{eqnarray}\label{B1}
&&\eta=4\epsilon\;\rightarrow\;F_G^\eta=F_L^\eta={\bf 1};\nonumber\\
&&\eta=4\epsilon+1\;\rightarrow\;F_G^\eta=F_G;\;\;\;F_L^\eta=F_L;\nonumber\\
&&\eta=4\epsilon+2\;\rightarrow\;F_G^\eta=F_L^\eta;\nonumber\\
&&\eta=4\epsilon+3\;\rightarrow\;F_G^\eta=F_G^\dagger;\;\;\;F_L^\eta=F_L^\dagger.
\end{eqnarray}
Therefore if ${\rm Tr}(F_G)={\rm Tr}(F_L)$ the $F_G, F_L$ are unitarily equivalent, and if ${\rm Tr}(F_G)\ne {\rm Tr}(F_L)$ the $F_G, F_L$ are unitarily inequivalent.

For an $n$-partite system with dimension $d^n$ we get ${\rm Tr}F_L=({\rm Tr}F)^n$ and using Eq.(\ref{R1}) we get
\begin{eqnarray}\label{B11}
&&d=4m+1\rightarrow\;{\rm Tr}F_L=1;\nonumber\\
&&d=4m+3\rightarrow\;{\rm Tr}F_L=i^n.
\end{eqnarray}
For the ${\rm Tr}F_G$ if $d=4m+1$, the $d^n=(4m+1)^n=4M_1 +1$ and we get
\begin{eqnarray}\label{B2}
d=4m+1\rightarrow\;{\rm Tr}F_G=1.
\end{eqnarray}
If $d=4m+3$
we consider two cases where $n$ is an even number and an odd number.
For even $n$ we get $d^n=(4m+3)^n=4M_2+1$ and for odd $n$ we find $d^n=(4m+3)^n=4M_3+3$. Therefore
\begin{eqnarray}\label{B3}
d=4m+3\;\;{\rm and}\;\;n={\rm even} \rightarrow\;{\rm Tr}F_G=1;\nonumber\\
d=4m+3\;\;{\rm and}\;\;n={\rm odd}\rightarrow\;{\rm Tr}F_G=i.
\end{eqnarray}
Comparison of Eq.(\ref{B11}) with Eqs(\ref{B2}),(\ref{B3}) proves the proposition.
\end{proof}

In the case of Eq.(\ref{59}) there exists a $d^n\times d^n$ unitary matrix $U$ such that $F_G=UF_LU^\dagger$, i.e., 
\begin{eqnarray}
\omega_{d^n}(\widehat i \widehat j)=\sum _{\widehat j,\widehat \ell}U(\widehat i,\widehat k)U(\widehat j,\widehat \ell)\omega_d(\ell_0k_0+...+\ell_{d-1}k_{n-1}).
\end{eqnarray}
So if instead of the basis $\ket{X;\widehat j}$ we choose the basis $U \ket{X;\widehat j}$ as position states, then the local Fourier transform with respect to the new basis is the global Fourier transform with respect to the old basis
$F_G=UF_LU^\dagger$.
So in this case the global Fourier transform is not a new concept.
However $U$ is in general a global transformation (it cannot be written as $U_1\otimes...\otimes U_n$) and for this reason there is some merit in the study of the global Fourier transform even in this case.

The case of Eq.(\ref{60}) where global and local Fourier transforms are unitarily inequivalent is clearly the most interesting one.
Then the global Fourier transform is a new concept.
In any case, the formalism below is the same for both cases in Eqs(\ref{59}), (\ref{60}).

\subsection{The global Fourier transform in terms of local Fourier transforms and applications in Fast Fourier transforms}\label{sec34}

The general idea of Fast Fourier transforms is to express the `large' Fourier transform in a large Hilbert space, as an `appropriate' combination of `small' Fourier transforms in smaller Hilbert spaces.
Performing the `small' Fourier transforms instead of the  `large' Fourier transform, is computationally beneficial.
The general formalism of this paper can be helpful in this direction.

As an example, we express in this section the global Fourier transform in terms of many local Fourier transforms.
This is similar to the Cooley-Tukey formalism in fast Fourier transforms\cite{B1,B2,B3}.
We only consider the special case $n=2$, and we do not discuss complexity issues.
But we point out that understanding of the relationship between global and local Fourier transforms and related phase space methods, can be useful in other areas like fast Fourier transforms.

 For the case $n=2$ we get
 \begin{eqnarray}
\omega_{d^2}({\widehat j}{\widehat k})=\omega _{d^2}(j_0k_0)\omega _d(j_0k_1+j_1k_0).
\end{eqnarray}

Let $\ket{s}$ be a quantum state in ${\mathfrak H}=H(d)\otimes H(d)$ and $s(k_0,k_1)=\bra{X;k_0,k_1}s\rangle$. Then
\begin{eqnarray}
\bra{X;j_0,j_1}F_G\ket{s}&=&\frac{1}{d}\sum _{k_0,k_1}\omega_{d^2}(\widehat j {\widehat k})s(k_0,k_1)\nonumber\\&=&
\frac{1}{d}\sum _{k_0}\omega_d(j_1k_0)\omega _{d^2}(j_0k_0)\sum _{k_1}\omega_d(j_0k_1)s(k_0,k_1)\nonumber\\&=&
\frac{1}{\sqrt d}\sum _{k_0}\omega_d(j_1k_0)\left[\omega _{d^2}(j_0k_0){\widetilde s}(k_0,j_0)\right ],
\end{eqnarray}
where
\begin{eqnarray}
{\widetilde s}(k_0,j_0)=\frac{1}{\sqrt d}\sum _{k_1}\omega_d(j_0k_1)s(k_0,k_1).
\end{eqnarray}
In this way the Fourier transform in a $d^2$-dimensional space reduces to two Fourier transforms in $d$-dimensional spaces.

\section{Global phase space methods}

\subsection{Global displacements in ${\mathbb Z}(d^n)\times {\mathbb Z}(d^n)$}
The phase space is  defined by the Fourier transform and for global Fourier transforms is ${\mathbb Z}(d^n)\times {\mathbb Z}(d^n)$.
Displacement operators in it are defined as

\begin{eqnarray}\label{55}
{\mathfrak X}_G(\widehat \beta) &=& \sum _{\widehat j}\omega _{d^n}(-\widehat j \widehat \beta)|P_G;\widehat j\rangle \bra{P_G; \widehat j}=\sum _{\widehat j}\ket{X; \widehat j+\widehat \beta}\bra{X;\widehat j},
\end{eqnarray}
and
\begin{eqnarray}\label{57}
{\mathfrak Z}_G(\widehat \alpha)& =&\sum _{\widehat j}|P_G; \widehat j+\widehat \alpha\rangle \bra{P_G;\widehat j}=\sum _{\widehat j}\omega _{d^n}(\widehat \alpha \widehat j)|X; \widehat j\rangle\bra{X;\widehat j}
=F_G{\mathfrak X}_G(\widehat \alpha)F_G^\dagger.
\end{eqnarray}
Addition in ${\mathbb Z}(d^n)$ is used in these two equations, in contrast to Eqs(\ref{A}), (\ref{XX}) where we have addition in $[{\mathbb Z}(d)]^n$.
The ${\mathfrak X}_L(\{\beta_r\})$ in Eq.(\ref{XX}) can also be written as 
\begin{eqnarray}\label{56}
{\mathfrak X}_L(\widehat \beta) &=& \sum _{\widehat j}\ket{X; \widehat {j+\beta}}\bra{X;\widehat j}.
\end{eqnarray}
We have explained that  $\widehat {j+\beta}\ne \widehat j+\widehat \beta$, and consequently Eqs(\ref{55}), (\ref{56}) are an
example of the difference between the local and global formalism. Also the ${\mathfrak Z}_L(\{\alpha _r\})$ in Eq.(\ref{A}) can also be written as
\begin{eqnarray}\label{58}
{\mathfrak Z}_L(\widehat \alpha)=
\sum _{j_r}\omega _{d}(\alpha _0j_0+...+\alpha_{n-1}j_{n-1})\ket{X; \widehat j}\bra{X;\widehat j}.
\end{eqnarray}
Comparison of Eqs(\ref{57}), (\ref{58}) again shows the difference between the local and global formalism.

Since ${\mathfrak Z}_G(\widehat \alpha){\mathfrak Z}_G(\widehat \gamma)={\mathfrak Z}_G(\widehat \alpha+\widehat \gamma)$, the ${\mathfrak Z}_G(\widehat \alpha)$ form a representation of 
${\mathbb Z}(d^n)$ as an additive group (which is not isomorphic to $[{\mathbb Z}(d)]^n$).
The same is true for the ${\mathfrak X}_G(\widehat \beta)$. 
Also
\begin{eqnarray}
&&{\mathfrak X}_G(\widehat \beta) {\mathfrak Z}_G(\widehat \alpha)= {\mathfrak Z}_G(\widehat \alpha){\mathfrak X}_G(\widehat \beta) \omega_{d^n}(-\widehat \alpha \widehat \beta);\nonumber\\
&&[{\mathfrak X}_G(\widehat \beta) ]^{d^n}=[{\mathfrak Z}_G(\widehat \alpha)]^{d^n}= {\bf 1}.
\end{eqnarray}
These relations should be compared and contrasted with Eq.(\ref{26}) for the local formalism.
Global displacement operators are defined as
\begin{eqnarray}
D_G(\widehat \alpha, \widehat \beta)={\mathfrak Z}_G(\widehat \alpha){\mathfrak X}_G(\widehat \beta)\omega _{d^n}(-2^{-1}\widehat \alpha \widehat \beta).
\end{eqnarray}
Here $2^{-1}=\frac{d^n+1}{2}$ is an element of ${\mathbb Z}(d^n)$.
The $D_G(\widehat \alpha, \widehat \beta)\omega  _{d^n}(\widehat \gamma)$ form a representation of the Heisenberg-Weyl group of displacements in the phase space ${\mathbb Z}(d^n)\times {\mathbb Z}(d^n)$.
We note that
\begin{eqnarray}
&&D_G(\widehat \alpha, \widehat \beta)\ket{X; \widehat j}=\omega_{d^n}(2^{-1}\widehat \alpha \widehat \beta +\widehat \alpha\widehat j)\ket{X; \widehat j+\widehat \beta};\nonumber\\
&&D_G(\widehat \alpha, \widehat \beta)\ket{P_G; \widehat j}=\omega_{d^n}(-2^{-1}\widehat \alpha \widehat \beta -\widehat \beta\widehat j)\ket{P_G; \widehat j+\widehat \alpha}.\nonumber\\
\end{eqnarray}
 Also
\begin{eqnarray}
D_G(\widehat \alpha, \widehat \beta)\ket{P_L; \widehat j}&=&\frac{1}{\sqrt{d^n}}\sum_{\widehat k}\omega _d(j_0k_0+...+j_{n-1}k_{n-1})D_G(\widehat \alpha, \widehat \beta)\ket{X; \widehat k}\nonumber\\
&=&\frac{1}{\sqrt{d^n}}\sum_{\widehat k}\omega _d(j_0k_0+...+j_{n-1}k_{n-1})\omega_{d^n}(2^{-1}\widehat \alpha \widehat \beta +\widehat \alpha\widehat k)\ket{X; \widehat k+\widehat \beta}.
\end{eqnarray}
These relations should be compared and contrasted to
\begin{eqnarray}
&&D_L(\widehat \alpha, \widehat \beta)\ket{X; \widehat j}=\omega_{d}[2^{-1}(\alpha_0\beta_0+...+\alpha_{n-1}\beta_{n-1})+(\alpha_0j_0+...+\alpha_{n-1}j_{n-1})]\ket{X; \widehat {j+ \beta}};\nonumber\\
&&D_L(\widehat \alpha, \widehat \beta)\ket{P_L; \widehat j}=\omega_{d}[-2^{-1}(\alpha_0\beta_0+...+\alpha_{n-1}\beta_{n-1})-(\beta_0j_0+...+\beta_{n-1}j_{n-1})]\ket{P_L; \widehat {j+ \alpha}}.
\end{eqnarray}
Also
\begin{eqnarray}
&&D_L(\widehat \alpha, \widehat \beta)\ket{P_G; \widehat j}=\frac{1}{\sqrt{d^n}}\sum _{\widehat k}\omega_{d^n}(\widehat j\widehat k)D_L(\widehat \alpha, \widehat \beta)\ket{X;\widehat k}\nonumber\\&&=
\frac{1}{\sqrt{d^n}}\sum _{\widehat k}\omega_{d^n}(\widehat j\widehat k)\omega_{d}[2^{-1}(\alpha_0\beta_0+...+\alpha_{n-1}\beta_{n-1})+(\alpha_0k_0+...+\alpha_{n-1}k_{n-1})]\ket{X; \widehat {k+ \beta}}.
\end{eqnarray}

The global parity operator (around the point  $(\widehat \gamma, \widehat \delta)$ in the phase space ${\mathbb Z}(d^n)\times {\mathbb Z}(d^n)$) is
\begin{eqnarray}
{\mathfrak P}_G(\widehat \gamma, \widehat \delta)=D_G(\widehat \gamma, \widehat \delta)F_G^2[D_G(\widehat \gamma, \widehat \delta)]^{\dagger};\;\;\;\left [{\mathfrak P}_G(\widehat \gamma, \widehat \delta)\right ]^2={\bf 1}.
\end{eqnarray}
In analogy to Eq.(\ref{11}) we find that the global parity operator is related to the global displacement operators through the Fourier transform 
\begin{eqnarray}\label{13}
&&{\mathfrak P}_G(\widehat\gamma,\widehat\delta)=\frac{1}{d^n}\sum_{\widehat \alpha, \widehat \beta}D_G(\widehat \alpha ,\widehat\beta)\omega_{d^n}(\widehat \beta \widehat \gamma-\widehat \alpha \widehat \delta);\nonumber\\
&&D_G(\widehat\alpha,\widehat\beta)=\frac{1}{d^n}\sum_{\widehat \gamma, \widehat \delta}{\mathfrak P}_G(\widehat\gamma,\widehat\delta)\omega_{d^n}(-\widehat\beta \widehat \gamma+\widehat \alpha \widehat \delta).
\end{eqnarray}

\begin{example}
We consider the case  $d=3$ and $n=2$. 
In this case the global Fourier transform is unitarily inequivalent to the local Fourier transform.
We work in the `periods ' of Eq.(\ref{47}).

Let $\ket{X;k_0,k_1}$ be the basis of position states.
 The globally Fourier transformed basis is
\begin{eqnarray}\label{dd}
\ket{P_G;j_0,j_1}
=\frac{1}{3}\sum _{k_0,k_1}\omega_{9}[j_0k_0+3(j_1k_0+j_0k_1)]\ket{X;k_0,k_1}.
\end{eqnarray}
The $j_r,k_r$ take the values $-1,0,1$.
The locally Fourier transformed basis is
\begin{eqnarray}
\ket{P_L;j_0,j_1}&=&\frac{1}{3}\sum _{k_0,k_1}\omega_{3}(j_0k_0+j_1k_1)\ket{X;k_0,k_1}.
\end{eqnarray}
Then 
\begin{eqnarray}
\langle P_L;\ell_0,\ell_1\ket{P_G;j_0,j_1}&=&\frac{1}{9}\{1+\omega_{9}(3j_0+9j_1)\omega _3(-\ell_1)+\omega_{9}(j_0+3j_1)\omega_3(-\ell_0)\nonumber\\&
+&\omega_{9}(4j_0+12j_1)\omega _3(-\ell_0-\ell_1)]\}.
\end{eqnarray}

We next consider the local displacement operator ${\mathfrak X}_L(-1,1)$ which acts on the 
states $\ket{X;1,0}$ and $\ket{P_L;1,0}$ as follows:
\begin{eqnarray}\label{S1}
&&{\mathfrak X}_L(-1,1)\ket{X;1,0}=\ket{X;0,1},
\end{eqnarray}
and 
\begin{eqnarray}\label{S2}
{\mathfrak X}_L(-1,1)\ket{P_L;1,0}=\omega_3(1)\ket{P_L;1,0}.
\end{eqnarray}
${\mathfrak X}_L(-1,1)$ acts on the state $\ket{P_G;1,0}$ as follows:
\begin{eqnarray}\label{S3}
{\mathfrak X}_L(-1,1)\ket{P_G;1,0}&=&{\mathfrak X}_L(-1,1)\sum_{j_1,j_0}\ket{X;j_0,j_1}\langle{X;j_0,j_1}\ket{P_G;1,0}\nonumber\\&=&
\frac{1}{3}\sum_{j_1,j_0}\omega_9(\widehat {j_0+3j_1})\ket{X;j_0-1,j_1+1}.
\end{eqnarray}
We also consider the corresponding global displacement operator ${\mathfrak X}_G(\widehat {-1+3\cdot 1})={\mathfrak X}_G(\widehat 2)$ which acts on the 
states $\ket{X;1,0}=\ket{X;\widehat 1}$ and $\ket{P_G;1,0}=\ket{P_G;\widehat 1}$ as follows:
\begin{eqnarray}\label{T1}
&&{\mathfrak X}_G(\widehat 2)\ket{X;\widehat 1}=\ket{X;\widehat 3}=\ket{X;0,1},
\end{eqnarray}
and 
\begin{eqnarray}\label{T3}
{\mathfrak X}_G(\widehat 2)\ket{P_G;\widehat 1}=\omega_9(-\widehat 2)\ket{P_G;\widehat 1}.
\end{eqnarray}
${\mathfrak X}_G(\widehat 5)$ acts on the state $\ket{P_L;1,0}$ as follows:
\begin{eqnarray}\label{T2}
&&{\mathfrak X}_G(\widehat 2)\ket{P_L;1,0}={\mathfrak X}_G(\widehat 2)\sum_{j_1,j_0}\ket{X;j_0,j_1}\langle{X;j_0,j_1}\ket{P_L;1,0}\nonumber\\&&=
\frac{1}{3}{\mathfrak X}_G(\widehat 2)\sum_{j_1,j_0}\ket{X;\widehat{j_0+3j_1}}\omega_3(j_0)=\frac{1}{3}\sum_{j_1,j_0}\omega_3(j_0)\ket{X;\widehat 2+\widehat{j_0+3j_1}}\nonumber\\&&=\frac{1}{3}\{
\omega_3(-1)[\ket{X;\widehat {-2}}+\ket{X;\widehat 1}+\ket{X;\widehat 4}]+[\ket{X;\widehat {-1}}+\ket{X;\widehat 2}+\ket{X;\widehat 5}]+
\omega _3(1)[\ket{X;\widehat 0}+\ket{X;\widehat 3}+\ket{X;\widehat 6}]\}.
\end{eqnarray}
Eqs. (\ref{S1}), (\ref{S2}), (\ref{S3}) involve local displacements and should be compared and contrasted to Eqs. (\ref{T1}), (\ref{T2}), (\ref{T3}) correspondingly, that involve global displacements.

\end{example}

\subsection{Local and global position and momentum operators and time evolution}\label{GGG}

We define local position and local momentum operators  for the $r$-component of the system as
\begin{eqnarray}\label{91}
&&{\cal X}_L^{(r)}={\bf 1}\otimes ...\otimes {\bf 1}\otimes {\cal X}\otimes {\bf 1}\otimes...\otimes {\bf 1};\;\;\;r=0,...,n-1;\nonumber\\
&&{\cal P}_L^{(r)}={\bf 1}\otimes ...\otimes {\bf 1}\otimes {\cal P}\otimes {\bf 1}\otimes...\otimes {\bf 1};\;\;\;F_L{\cal P}_LF_L^\dagger=-{\cal X}_L.
\end{eqnarray}
The ${\cal X}, {\cal P}$ have been defined in Eq.(\ref{9}).
We can also define global position and global momentum operators as
\begin{eqnarray}\label{92}
{\cal X}_G=-i\frac{d^n}{2\pi}\log[{\mathfrak Z}_G(\widehat 1)];\;\;\;{\cal P}_G=i\frac{d^n}{2\pi}\log[{\mathfrak X}_G(\widehat 1)]
;\;\;\;F_G{\cal P}_GF_G^\dagger=-{\cal X}_G.
\end{eqnarray}
They all are $d^n\times d^n$ matrices which can be interpreted as position and momentum operators. 
In a multipartite system with weak interaction between the various parties, it can be argued that the local variables ${\cal X}_L^{(r)}, {\cal P}_L^{(r)}$ are more physical operators and the Hamiltonian should be expressed in terms of them.
But in the case of strong interactions between the parties, the global variables ${\cal X}_G, {\cal P}_G$ might be better for a holistic simple description of the physical Hamiltonian with a good approximation.

\begin{example}
We consider the case $d=3$, $n=2$ and the quantum state
\begin{eqnarray}\label{91}
\ket{s}=\frac{1}{\sqrt {84}}(\ket{X;1}+2\ket{X;0}-3\ket{X;-1})\otimes(i\ket{X;1}+\ket{X;0}-2i\ket{X;-1}).
\end{eqnarray}
We also consider time evolution with the Hamiltonians
\begin{eqnarray}
&&h_1=\frac{1}{2}[{\cal P}_G^2+{\cal X}_G^2];\nonumber\\
&&h_2=\frac{1}{2}({\cal P}^2\otimes {\bf 1})+\frac{1}{2}({\cal X}^2\otimes {\bf 1})+\frac{1}{2}({\bf 1}\otimes {\cal P}^2)+\frac{1}{2}({\bf 1}\otimes {\cal X}^2)+{\cal X}\otimes{\cal X}.
\end{eqnarray}
The first uses the global momentum and position, and the second uses the local momenta and positions.
 Here (in the position basis)
 \begin{eqnarray}\label{AB}
&&{\cal X}=\ket{X;1}\bra{X;1}-\ket{X;-1}\bra{X;-1};\;\;\;{\cal P}=-F^\dagger {\cal X}F
 \end{eqnarray}
 Using both notations ${\cal X}_G$ and ${\cal P}_G$ are
 \begin{eqnarray}\label{AC}
&&{\cal X}_G=\sum_{a=-1}^1\sum_{b=-1}^1(a+3b)\ket{X;a,b}\bra{X;a,b}=\sum_{a=-1}^1\sum_{b=-1}^1(a+3b)\ket{X;\widehat{a+3b}}\bra{X;\widehat{a+3b}}\nonumber\\&&{\cal P}_G=-F_G^\dagger {\cal X}_GF_G.
 \end{eqnarray}

At time $t=1$ the state becomes
\begin{eqnarray}
\exp(ith_1)\ket{s}=\sum_{a=-1}^1\sum_{b=-1}^1\lambda_1(a,b)\ket{X;a,b}
;\;\;\;\exp(ith_2)\ket{s}=\sum_{a=-1}^1\sum_{b=-1}^1\lambda_2(a,b)\ket{X;a,b}  
\end{eqnarray}
where
\begin{eqnarray}
\lambda_1(-1,-1) =-0.2337 - 0.3556i;\;\; \lambda_1(-1,0)=0.0259 - 0.1315i;\;\;
\lambda_1(-1,1) = 0.3254 + 0.4898i;\nonumber\\ \lambda_1(0,-1)=0.1160 + 0.0430i;\;\;
\lambda_1(0,0) = 0.2836 - 0.0628i;\;\; \lambda_1(0,1)=-0.3090 + 0.1367i;\nonumber\\
\lambda_1(1,-1) = 0.0671 + 0.0943i;\;\; \lambda_1(1,0)=-0.3539 - 0.0845i;\;\;\lambda_1(1,1)=0.3014 - 0.0675i;
\end{eqnarray}
and
\begin{eqnarray}
\lambda_2(-1,-1) = -0.2911 - 0.7197i;\;\; \lambda_2(-1,0)=-0.1491 - 0.2668i;\;\;
\lambda_2(-1,1) = 0.1502 - 0.0831i;\nonumber\\ \lambda_2(0,-1)=0.3700 - 0.0854i;\;\;
\lambda_2(0,0) = 0.1218 + 0.1319i;\;\; \lambda_2(0,1)=-0.2208 - 0.0247i;\nonumber\\
\lambda_2(1,-1) = 0.0872 - 0.1268i;\;\; \lambda_2(1,0)=-0.0032 + 0.0910i;\;\;\lambda_2(1,1)=-0.0340 - 0.1243i.
\end{eqnarray}

\end{example}

\subsection{Global Wigner and global Weyl functions in ${\mathbb Z}(d^n)\times {\mathbb Z}(d^n)$}

If $\rho$ is a density matrix,  we define the global Wigner function $W_G(\widehat\gamma, \widehat\delta|\rho)$ and the global Weyl function $\widetilde W_G( \widehat\alpha, \widehat\beta|\rho)$ as:
\begin{eqnarray}\label{WG}
W_G(\widehat\gamma,\widehat\delta|\rho)={\rm Tr}[ \rho {\mathfrak P}_G(\widehat\gamma,\widehat \delta)];\;\;\;{\widetilde W_G}(\widehat\alpha ,\widehat\beta|\rho)={\rm Tr}[ \rho D_G(\widehat\alpha,\widehat \beta)].
\end{eqnarray}

From Eq.(\ref{13}) it follows that they are related to each other through the Fourier transform:
\begin{eqnarray}
&&W_G(\widehat\gamma,\widehat\delta|\rho)=\frac{1}{d^n}\sum_{\widehat\alpha,\widehat\beta}{\widetilde W_G}(\widehat\alpha,\widehat\beta|\rho)\omega_{d^n}(\widehat\beta \widehat\gamma-\widehat \alpha \widehat\delta);\nonumber\\
&&{\widetilde W_G}(\widehat\alpha,\widehat\beta|\rho)=\frac{1}{d^n}\sum_{\widehat\gamma, \widehat\delta}W_G(\widehat\gamma,\widehat\delta|\rho)\omega_{d^n}(-\widehat\beta\widehat\gamma+\widehat\alpha \widehat\delta).
\end{eqnarray}

The following marginal properties of the Wigner function follow immediately from Eq.(\ref{A1}) (for odd values of the dimension $d$):
\begin{eqnarray}\label{C}
&&\frac{1}{d^n}\sum _{\{\gamma_r\}} W_L(\{\gamma_r,\delta_r\}|\rho)=\frac{1}{d^n}\sum _{\widehat \gamma} W_G(\widehat \gamma,\widehat \delta |\rho)=\bra{X;\delta_0,...,\delta_{n-1}}\rho\ket{X;\delta_0,...,\delta_{n-1}}=
\bra{X;\widehat \delta}\rho\ket{X;\widehat \delta};
\nonumber\\
&&\frac{1}{d^n}\sum _{\{\delta_r\}} W_L(\{\gamma_r,\delta_r\}|\rho)=\bra{P_L;\gamma_0,...,\gamma_{n-1}}\rho\ket{P_L;\gamma_0,...,\gamma_{n-1}};\nonumber\\
&&\frac{1}{d^n}\sum _{\widehat \delta} W_G(\widehat \gamma,\widehat \delta|\rho)=\bra{P_G;\widehat \gamma}\rho\ket{P_G;\widehat \gamma};\nonumber\\
&&\frac{1}{d^n}\sum _{\{\gamma_r,\delta_r\}} W_L(\{\gamma_r,\delta_r\}|\rho)=\frac{1}{d^n}\sum _{\widehat \gamma,\widehat \delta} W_G(\widehat \gamma,\widehat \delta|\rho)=1.
\end{eqnarray}
We have already emphasized that in both the `local formalism' and the `global formalism' the position states are the same and the momentum states are different.
Consequently $\bra{P_L;\gamma_0,...,\gamma_{n-1}}\rho\ket{P_L;\gamma_0,...,\gamma_{n-1}}$ is different from 
$\bra{P_G;\widehat \gamma}\rho\ket{P_G;\widehat \gamma}$ and the marginal properties in the second and third of these equations are different.

\subsection{The difference between the local and global Wigner functions}

We first consider states for which the local Wigner function is the same as the global Wigner function.
\begin{proposition}\label{pro22}
We consider the following separable density matrix that contains only diagonal elements with respect to the basis of position states:
\begin{eqnarray}\label{dg}
\sigma=\sum _{\widehat j}p(\widehat j)\ket{X;\widehat j}\bra{X;\widehat j};\;\;\;\sum _{\widehat j}p(\widehat j)=1.
\end{eqnarray}
Here the $p(\widehat j)$ are probabilities.
In this case the local and global Wigner functions are equal to each other, they are non-negative and they do not depend on $\widehat \alpha$:
\begin{eqnarray}
W_G(\widehat\alpha,\widehat\beta|\sigma)=W_L(\{\alpha_r,\beta_r\}|\sigma)=p(\widehat {\beta}).
\end{eqnarray} 
\end{proposition}
\begin{proof}
For the position states 
\begin{eqnarray}
\sigma_0(\widehat j)=\ket{X;\widehat j}\bra{X;\widehat j}=\ket{X;j_0,...,j_{n-1}}\bra{X;j_0,...,j_{n-1}},
\end{eqnarray}
we get
\begin{eqnarray}\label{PL}
W_G(\widehat\alpha,\widehat\beta|\sigma_0(\widehat j))=W_L(\{\alpha_r,\beta_r\}|\sigma_0 (\widehat j))=\delta(\widehat {j},\widehat {\beta}).
\end{eqnarray} 
Indeed
\begin{eqnarray}
W_G(\widehat\alpha,\widehat\beta|\sigma_0(\widehat j))&=&{\rm Tr}\left [{\mathfrak P}_G(\widehat \alpha, \widehat \beta)\ket{X;\widehat j}\bra{X;\widehat j}\right ]\nonumber\\&=&
{\rm Tr}\left \{F_G^2[D_G(\widehat \alpha, \widehat \beta)]^{\dagger}\ket{X;\widehat{j}}\bra{X;\widehat {j}}D_G(\widehat \alpha, \widehat \beta)\right \}\nonumber\\&=&
{\rm Tr}\left [F_G^2\ket{X;\widehat {j}-\widehat{\beta}}\bra{X;\widehat {j}-\widehat {\beta}}\right ]=\delta(\widehat {j},\widehat {\beta}),
\end{eqnarray} 
and  also
\begin{eqnarray}
W_L(\{\alpha_r,\beta_r\}|\sigma_0(\widehat j))&=&{\rm Tr}[ {\mathfrak P}_L(\{\alpha_r,\beta_r\})\ket{X;j_0,...,j_{n-1}}\bra{X;j_0,...,j_{n-1}}]\nonumber\\&=&
{\rm Tr}\left \{F_L^2[D_L(\{\alpha_r,\beta_r\})]^{\dagger}\ket{X;j_0,...,j_{n-1}}\bra{X;j_0,...,j_{n-1}}D_L(\{\alpha_r,\beta_r\})\right \}\nonumber\\&=&
{\rm Tr}\left [F_L^2\ket{X;j_0-\beta _0,...,j_{n-1}-\beta_{n-1}}\bra{X;j_0-\beta _0,...,j_{n-1}-\beta_{n-1}}\right ]\nonumber\\&=&\delta (j_0,\beta _0)...\delta(j_{n-1},\beta_{n-1}).
\end{eqnarray} 
This proves Eq.(\ref{PL}). Then
\begin{eqnarray}
W_G(\widehat\alpha,\widehat\beta|\sigma)=\sum _{\widehat j}p(\widehat j)W_G(\widehat\alpha,\widehat\beta|\sigma_0(\widehat j))=\sum _{\widehat j}p(\widehat j)\delta(\widehat {j},\widehat {\beta})=p(\widehat {\beta}).
\end{eqnarray} 
and also
\begin{eqnarray}
W_L(\{\alpha_r,\beta_r\}|\sigma)&=&\sum _{j_0,...,j_{n-1}}p(j_0,...,j_{n-1})W_L(\{\alpha_r,\beta_r\}|\sigma_0(\widehat j))\nonumber\\&=&\sum _{j_0,...,j_{n-1}}p(j_0,...,j_{n-1})\delta (j_0,\beta _0)...\delta(j_{n-1},\beta_{n-1})=p(\widehat {\beta}).
\end{eqnarray}

\end{proof}

An arbitrary density matrix $\rho$ can be written in the basis of position states, as the sum of a separable density matrix $\sigma (\rho)$ that contains the $d^n$ diagonal elements (as in Eq.(\ref{dg})), 
and a Hermitian matrix $\tau(\rho)$ with trace zero that contains the $d^{2n}-d^n$ off-diagonal elements:
\begin{eqnarray}
\rho=\sigma(\rho)+\tau(\rho);\;\;\;{\rm Tr}[\tau(\rho)]=0.
\end{eqnarray}
$\tau(\rho)$ is not a density matrix but using Eqs(\ref{WL}), (\ref{WG}) we can define `Wigner-like' functions for it.
Then the Wigner function is written as a sum of two terms that correspond to the diagonal and off-diagonal part:
\begin{eqnarray}
W_L(\{\alpha_r,\beta_r\}|\rho)=W_L(\{\alpha_r,\beta_r\}|\sigma(\rho))+A_L(\{\alpha_r,\beta_r\}|\tau(\rho));\;\;\;
A_L(\{\alpha_r,\beta_r\}|\tau(\rho))={\rm Tr}[ \tau(\rho) {\mathfrak P}_L(\{\alpha_r,\beta_r\})];
\end{eqnarray} 
and 
\begin{eqnarray}
W_G(\widehat\alpha,\widehat\beta|\rho)=W_G(\widehat\alpha,\widehat\beta|\sigma(\rho))+A_G(\widehat \alpha, \widehat\beta|\tau(\rho));\;\;\;
A_G(\widehat\alpha,\widehat \beta|\tau(\rho))={\rm Tr}[ \tau(\rho) {\mathfrak P}_G(\widehat \alpha,\widehat \beta)].
\end{eqnarray} 
Then
\begin{eqnarray}
W_L(\{\alpha_r,\beta_r\}|\rho)-W_G(\widehat\alpha,\widehat\beta|\rho)=A_L(\{\alpha_r,\beta_r\}|\tau(\rho))-A_G(\widehat\alpha,\widehat\beta|\tau(\rho)).
\end{eqnarray}
The difference between local and global Wigner functions, is related only to the off-diagonal elements of the density matrix (with respect to the position basis).

\begin{proposition}\label{pro200}
We consider the following separable density matrices
\begin{eqnarray}
q_L=\sum _{\widehat j}p(\widehat j)\ket{P_L;\widehat j}\bra{P_L;\widehat j};\;\;\;q_G=\sum _{\widehat j}p(\widehat j)\ket{P_G;\widehat j}\bra{P_G;\widehat j};\;\;\;
F_L^\dagger q_LF_L=F_G^\dagger q_GF_G;\;\;\;\sum _{\widehat j}p(\widehat j)=1.
\end{eqnarray}
Here the $p(\widehat j)$ are probabilities. Then
\begin{eqnarray}
W_G(\widehat\alpha,\widehat\beta|q_G)=W_L(\{\alpha_r,\beta_r\}|q_L)=p(\widehat {\alpha}).
\end{eqnarray} 
\end{proposition}
\begin{proof}
We first consider the density matrices
\begin{eqnarray}
q_{L0}(\widehat j)=\ket{P_L;\widehat j}\bra{P_L;\widehat j}=\ket{P_L;j_0,...,j_{n-1}}\bra{P_L;j_0,...,j_{n-1}};\;\;\;q_{G0}(\widehat j)=\ket{P_G;\widehat j}\bra{P_G;\widehat j}.
\end{eqnarray}
and prove that
\begin{eqnarray}\label{PLL}
W_G(\widehat\alpha,\widehat\beta |q_{G0}(\widehat j))=W_L(\{\alpha_r,\beta_r\}|q_{L0} (\widehat j))=\delta(\widehat {j},\widehat {\alpha}).
\end{eqnarray} 
Indeed
\begin{eqnarray}
W_G(\widehat\alpha,\widehat\beta|q_{G0}(\widehat j))&=&{\rm Tr}\left [{\mathfrak P}_G(\widehat \alpha, \widehat \beta)\ket{P_G;\widehat j}\bra{P_G;\widehat j}\right ]\nonumber\\&=&
{\rm Tr}\left \{F_G^2[D_G(\widehat \alpha, \widehat \beta)]^{\dagger}\ket{P_G;\widehat{j}}\bra{P_G;\widehat {j}}D_G(\widehat \alpha, \widehat \beta)\right \}\nonumber\\&=&
{\rm Tr}\left [F_G^2\ket{P_G;\widehat {j}-\widehat{\alpha}}\bra{P_G;\widehat {j}-\widehat {\alpha}}\right ]=\delta(\widehat {j},\widehat {\alpha}),
\end{eqnarray} 
and
\begin{eqnarray}
W_L(\{\alpha_r,\beta_r\}|q_{L0}(\widehat j))&=&{\rm Tr}[ {\mathfrak P}_L(\{\alpha_r,\beta_r\})\ket{P_L;j_0,...,j_{n-1}}\bra{P_L;j_0,...,j_{n-1}}]\nonumber\\&=&
{\rm Tr}\left \{F_L^2[D_L(\{\alpha_r,\beta_r\})]^{\dagger}\ket{P_L;j_0,...,j_{n-1}}\bra{P_L;j_0,...,j_{n-1}}D_L(\{\alpha_r,\beta_r\})\right \}\nonumber\\&=&
{\rm Tr}\left [F_L^2\ket{P_L;j_0-\alpha _0,...,j_{n-1}-\alpha_{n-1}}\bra{P_L;j_0-\alpha _0,...,j_{n-1}-\alpha_{n-1}}\right ]\nonumber\\&=&\delta (j_0,\alpha _0)...\delta(j_{n-1},\alpha_{n-1}).
\end{eqnarray} 
This proves Eq.(\ref{PLL}). Then
\begin{eqnarray}
W_G(\widehat\alpha,\widehat\beta|q_G)=\sum _{\widehat j}p(\widehat j)W_G(\widehat\alpha,\widehat\beta|q_{G0}(\widehat j))=\sum _{\widehat j}p(\widehat j)\delta(\widehat {j},\widehat {\alpha})=p(\widehat {\alpha}).
\end{eqnarray} 
and also
\begin{eqnarray}
W_L(\{\alpha_r,\beta_r\}|q_L)&=&\sum _{j_0,...,j_{n-1}}p(\widehat j)W_L(\{\alpha_r,\beta_r\}|q_{L0}(\widehat j))\nonumber\\&=&\sum _{j_0,...,j_{n-1}}p(j_0,...,j_{n-1})\delta (j_0,\alpha _0)...\delta(j_{n-1},\alpha_{n-1})=p(\widehat {\alpha}).
\end{eqnarray}

\end{proof}

\begin{example}
We consider the density matrices
\begin{eqnarray}\label{rholg}
\rho_L=\ket{P_L;\widehat j}\bra{P_L;\widehat j};\;\;\;\rho_G=\ket{P_G;\widehat j}\bra{P_G;\widehat j};\;\;\;\rho_0=\frac{1}{d^n}{\bf 1}_{d^n}.
\end{eqnarray} 
Then 
\begin{eqnarray}
&&\sigma(\rho_L)=\sigma(\rho_G)=\sigma(\rho_0)=\rho_0=\frac{1}{d^n}{\bf 1}_{d^n};\nonumber\\
&&\tau(\rho_L)=\ket{P_L;\widehat j}\bra{P_L;\widehat j}-\frac{1}{d^n}{\bf 1}_{d^n};\;\;\;\tau(\rho_G)=\ket{P_L;\widehat j}\bra{P_L;\widehat j}-\frac{1}{d^n}{\bf 1}_{d^n};\;\;\;\tau(\rho_0)=0.
\end{eqnarray} 
In this case 
\begin{eqnarray}
W_G(\widehat\alpha,\widehat\beta|\rho_0)=W_L(\{\alpha_r,\beta_r\}|\rho_0)=\frac{1}{d^{n}}.
\end{eqnarray} 
From proposition \ref{pro200} it follows that
\begin{eqnarray}
W_G(\widehat\alpha,\widehat\beta|\rho_G)=W_L(\widehat\alpha,\widehat\beta|\rho_L)=\delta(\widehat {j},\widehat {\alpha}).
\end{eqnarray}
We next calculate numerically the  $W_L(\widehat\alpha,\widehat\beta|\rho_G)$, $W_G(\widehat\alpha,\widehat\beta|\rho_L)$, for an example. We consider the case $d=3$ and $n=2$, and the density matrices $\rho_L, \rho_G$ with $\widehat j=4$ (which is an element of ${\mathbb Z}(9)$).
Results for the  $W_L(\widehat\alpha,\widehat\beta|\rho_G)$, $W_G(\widehat\alpha,\widehat\beta|\rho_L)$, given in tables \ref{t1}, \ref{t2}  correspondingly.

\end{example}

\subsection{The $R_L, R_G$ matrices: indicators of classical and quantum correlations}\label{sec35}
In this section we compare quantities for $\rho$ with the corresponding quantities for $\mathfrak R (\rho)$ (in Eq.(\ref{1010})).
\begin{definition}
If $\rho$ is a density matrix,
$R_L$, $\widetilde R_L$ are $d^n\times d^n$ matrices with elements
\begin{eqnarray}
&&R_L(\{\gamma_r,\delta_r\}|\rho)=W_L(\{\gamma_r,\delta_r\}|\rho)-W_L(\{\gamma_r,\delta_r\}|{\mathfrak R}(\rho));\nonumber\\
&&\widetilde R_L(\{\alpha_r,\beta_r\}|\rho)={{\widetilde W_L}(\{\alpha_r,\beta_r\}|\rho)}-{{\widetilde W_L}(\{\alpha_r,\beta_r\}|{\mathfrak R}(\rho))}.
\end{eqnarray} 
Also $R_G$, $\widetilde R_G$ are $d^n\times d^n$ matrices with elements 
\begin{eqnarray}
&&R_G((\widehat \gamma,\widehat \delta|\rho)=W_G(\widehat \gamma,\widehat \delta|\rho)-W_G(\widehat \gamma,\widehat \delta|\mathfrak R(\rho));\nonumber\\
&&\widetilde R_G(\widehat \alpha,\widehat \beta|\rho)={\widetilde W_G}(\widehat\alpha,\widehat\beta|\rho)-{\widetilde W_G}(\widehat\alpha,\widehat\beta|\mathfrak R(\rho)).
\end{eqnarray} 
\end{definition}
\begin{proposition}
\begin{itemize}
\mbox{}
\item[(1)]
For factorisable density matrices $R_L=\widetilde R_L=R_G=\widetilde R_G=0$.
\item[(2)]
The $R_L$ and $\widetilde R_L$ are related through a local Fourier transform:
\begin{eqnarray}
&&R_L(\{\gamma_r,\delta_r\}|\rho)=\frac{1}{d^n}\sum_{\{\alpha_r,\beta_r\}}{\widetilde R}_L(\{\alpha_r,\beta_r\}|\rho)\omega_d\left [\sum_{r=0}^{n-1}(\beta _r\gamma_r-\alpha _r\delta_r)\right ];\nonumber\\
&&{\widetilde R}_L(\{\alpha_r,\beta_r\}|\rho)=\frac{1}{d^n}\sum_{\{\gamma_r,\delta_r\}}R_L(\{\gamma_r,\delta_r\}|\rho)\omega_d\left [\sum_{r=0}^{n-1}(-\beta _r\gamma_r+\alpha _r\delta_r)\right ].
\end{eqnarray}

\item[(3)]
The $R_G$ and $\widetilde R_G$ are related through a global Fourier transform:
\begin{eqnarray}
&&R_G(\widehat\gamma,\widehat\delta|\rho)=\frac{1}{d^n}\sum_{\widehat \alpha, \widehat \beta}\widetilde R_G(\widehat \alpha ,\widehat\beta|\rho)\omega_{d^n}(\widehat \beta \widehat \gamma-\widehat \alpha \widehat \delta);\nonumber\\
&&\widetilde R_G(\widehat\alpha,\widehat\beta|\rho)=\frac{1}{d^n}\sum_{\widehat \gamma, \widehat \delta}R_G(\widehat\gamma,\widehat\delta|\rho)\omega_{d^n}(-\widehat\beta \widehat \gamma+\widehat \alpha \widehat \delta).
\end{eqnarray}
\item[(4)]
The following are marginal properties:
\begin{eqnarray}\label{D}
&&\frac{1}{d^n}\sum _{\{\gamma_r\}} R_L(\{\gamma_r,\delta_r\}|\rho)=\frac{1}{d^n}\sum _{\widehat \gamma} R_G(\widehat \gamma,\widehat \delta|\rho)={\cal C}(X;\delta_0,...,\delta_{n-1})={\cal C}(X;\widehat \delta);\nonumber\\
&&\frac{1}{d^n}\sum _{\{\delta_r\}} R_L(\{\gamma_r,\delta_r\}|\rho)={\cal C}(P_L;\gamma_0,...,\gamma_{n-1});\nonumber\\
&&\frac{1}{d^n}\sum _{\widehat \delta} R_G(\widehat \gamma, \widehat \delta|\rho)={\cal E}(P_G;\widehat \gamma);\nonumber\\
&&\frac{1}{d^n}\sum _{\{\gamma_r,\delta_r\}} R_L(\{\gamma_r,\delta_r\}|\rho)=\frac{1}{d^n}\sum _{\widehat \gamma,\widehat \delta} R_G(\widehat \gamma,\widehat \delta|\rho)=0.
\end{eqnarray}

\end{itemize}
\end{proposition}
\begin{proof}
\begin{itemize}
\item[(1)]
For factorisable density matrices ${\mathfrak R}(\rho)=\rho$ and then $R_L=\widetilde R_L=R_G=\widetilde R_G=0$.

\item[(2)]
We prove this using Eq.(\ref{12}) with both $\rho$ and ${\mathfrak R}(\rho)$.
\item[(3)]
We prove this using Eq.(\ref{58}) with both $\rho$ and ${\mathfrak R}(\rho)$.
\item[(4)]
We prove this using Eq.(\ref{C}) with both $\rho$ and ${\mathfrak R}(\rho)$.

\end{itemize}
\end{proof}
The matrices $R_\rho$ and $\widetilde R_\rho$ indicate the existence of both classical and quantum correlations.

\section{Examples}
In the examples below we take  $d=3$ and $n=2$. 
In this case the global Fourier transform is unitarily inequivalent to the local Fourier transform.
We work in the `periods ' of Eq.(\ref{47}).

We consider the density matrix
\begin{eqnarray}\label{68}
\rho=\ket{s}\bra{s};\;\;\;\ket{s}=\frac{1}{\sqrt{3}}\ket{X;0,1}+\frac{1}{2}\ket{X;1,-1}+\sqrt{\frac{5}{12}}\ket{X;-1,0}.
\end{eqnarray}
The state described by $\rho$ is entangled.
In this case the reduced density matrices are
\begin{eqnarray}
&&\breve \rho_0=\frac{1}{3}\ket{X;0}\bra{X;0}+\frac{1}{4}\ket{X;1}\bra{X;1}+\frac{5}{12}\ket{X;-1}\bra{X;-1};\nonumber\\
&&\breve \rho_1=\frac{5}{12}\ket{X;0}\bra{X;0}+\frac{1}{3}\ket{X;1}\bra{X;1}+\frac{1}{4}\ket{X;-1}\bra{X;-1}.
\end{eqnarray}
In tables \ref{t3},\ref{t4},\ref{t5} and \ref{t6} we present  the local Wigner function $W_L(\gamma_0,\gamma_1; \delta_0, \delta _1)$,
the local Weyl function ${\widetilde W}_L(\alpha_0,\alpha_1; \beta _0;\beta_1)$, 
and the matrices $R_L(\gamma_0,\gamma_1; \delta_0, \delta _1)$ and ${\widetilde R}_L(\alpha_0,\alpha_1; \beta _0;\beta_1)$
for the density matrix $\rho$ in Eq.(\ref{68}).

The correlations in Eqs.(\ref{BB1}),(\ref{BB2}) are
\begin{equation}
{\cal C}_\rho(X;\delta_0,\delta_1)=
  \begin{bmatrix}
    -0.1042 & 0.2431 & -0.1389  \\
    -0.0833 & -0.1389 & 0.2222 \\
    0.1875  & -0.1042 & -0.0833
  \end{bmatrix};\;\;\;
  {\cal C}_\rho(P_L;\gamma_0,\gamma_1)=
  \begin{bmatrix}
    -0.1093 & -0.1093 & 0.2187  \\
    -0.1093 & 0.2187 &  -0.1093\\
    0.2187  & -0.1093 & -0.1093
  \end{bmatrix}.\;\;\;
   \label{eq:myeqn}
\end{equation}
We easily confirm that Eqs(\ref{C}) hold for the local Wigner and Weyl function.

For the global formalism in ${\mathbb Z}(9)$ we rewrite $\rho$ as
\begin{eqnarray}\label{68AA}
\rho=\ket{s}\bra{s};\;\;\;\ket{s}=\frac{1}{\sqrt{3}}\ket{X;\widehat 3}+\frac{1}{2}\ket{X;\widehat {-2}}+\sqrt{\frac{5}{12}}\ket{X;\widehat {-1}}.
\end{eqnarray}
In tables \ref{t7},\ref{t8}, \ref{t9} and \ref{t10} we present the global Wigner function $W_G(\widehat \gamma,\widehat \delta |\rho)$, the global Weyl function ${\widetilde W}_G(\widehat \alpha,\widehat \beta |\rho)$
and the matrices $R_G(\widehat{\gamma}, \widehat{\delta})$ and $\widetilde{R_G}(\widehat{\alpha}, \widehat{\beta})$ for the density matrix $\rho$ in Eq.(\ref{68}).

The correlations in Eq.(\ref{C1}) are 
\begin{eqnarray}
&&{\cal C}_\rho(X;\widehat \delta)={\cal C}_\rho(X;\delta_0,\delta_1);\nonumber\\
&&{\cal E}_\rho(P_G;\widehat \gamma)=\begin{bmatrix}
    -0.0419 &-0.1093 & 0.1250  &
    -0.0832 & 0.2187 & -0.0832 &
    0.1250  & -0.1093 & -0.0419
  \end{bmatrix}^T.
  \end{eqnarray}
We easily confirm that Eqs(\ref{C}) hold for the global Wigner and Weyl function.

In general there is no simple relation that links the local with the global quantities.
We see this by comparing the expectation values of the local observables ${\cal X}\otimes {\bf 1}$, ${\bf 1}\otimes {\cal X}$, ${\cal X}\otimes {\cal X}$, ${\cal P}\otimes {\bf 1}$,  
 ${\bf 1}\otimes {\cal P}$, ${\cal P}\otimes {\cal P}$ for the density matrix $\rho$ in Eq.(\ref{68}), 
 with the expectation values of the global observables ${\cal X}_G$,  ${\cal P}_G$ for the same density matrix (written in the `global language' in Eq.(\ref{68AA})):
 \begin{eqnarray}
&&{\rm Tr}[\rho ({\cal X}\otimes {\bf 1})]=-0.1667;\;\;\;{\rm Tr}[\rho ({\bf 1}\otimes {\cal X})]=0.0833;\;\;\;{\rm Tr}[\rho ({\cal X}\otimes {\cal X})]=-0.25;\;\;\;{\rm Tr}(\rho {\cal X}_G)=0.0833\nonumber\\
&&{\rm Tr}[\rho ({\cal P}\otimes {\bf 1})]=0;\;\;\;{\rm Tr}[\rho ({\bf 1}\otimes {\cal P})]=0;\;\;\;{\rm Tr}[\rho ({\cal P}\otimes {\cal P})]=-0.6561;\;\;\;{\rm Tr}(\rho {\cal P}_G)=0.
 \end{eqnarray}
The ${\cal X}$, ${\cal P}$, ${\cal X}_G$, ${\cal P}_G$, have been given in Eqs(\ref{AB}), (\ref{AC}). 

The results for the local observables are different from the global observables.
For strongly correlated systems global quantities might be physically more relevant.

We note that an observable can be written in both the local and global formalism (using the map in Eq.(\ref{16})).
For example, for the above system we consider the observable (Hermitian operator)
 \begin{eqnarray}
{\cal O}&=&a\ket{X;0,1}\bra{X;0,1}+b\ket{X;1,-1}\bra{X;1,-1}+c\ket{X;-1,0}\bra{X;-1,0}\nonumber\\&+&d\ket{X;0,1}\bra{X;1,-1}+d^*\ket{X;1,-1}\bra{X;0,1};\;\;\;
a,b,c\in{\mathbb R};\;\;\;d\in{\mathbb C},
 \end{eqnarray}
which can also be written as
 \begin{eqnarray}
{\cal O}&=&a\ket{X;\widehat 3}\bra{X;\widehat 3}+b\ket{X;\widehat {-2}}\bra{X;\widehat {-2}}+c\ket{X;\widehat {-1}}\bra{X;\widehat {-1}}\nonumber\\&+&d\ket{X;\widehat 3}\bra{X;\widehat {-2}}+d^*\ket{X;\widehat {-2}}\bra{X;\widehat 3}.
 \end{eqnarray}

Important physical quantities like the position can be defined locally like 
${\cal X}\otimes {\bf 1}$, ${\bf 1}\otimes {\cal X}$, ${\cal X}\otimes {\cal X}$, or globally as ${\cal X}_G$ (defined in Eq.(\ref{92})).
The same is true  for local and global momenta.
For strongly correlated systems the identity of each component becomes weak, and global quantities might be physically more appropriate for the description of these systems.

\section{Discussion}

In this paper we introduced local and global Fourier transforms and related phase space methods for multipartite systems.
The multipartite system consists of $n$ components, each of which is described with variables in ${\mathbb Z}(d)$ and with a $d$-dimensional Hilbert space $H(d)$.
In the global formalism we take a holistic view of the system and describe it with variables in $[{\mathbb Z}(d^n)]$ and the $d^n$-dimensional Hilbert space ${\mathfrak H}$. 
Even if the various components of the system are located far from each other, in the case of strong interactions and strong correlations between them they might loose their individual identity.
 In this case a holistic approach that uses global quantities, might be more appropriate.

In the local formalism the phase space is $[{\mathbb Z}(d)\times {\mathbb Z}(d)]^n$, and in the global formalism $[{\mathbb Z}(d^n)]\times [{\mathbb Z}(d^n)]$.
We have explained that although the map in Eq.(\ref{16}) is bijective, the ring $[{\mathbb Z}(d)]^n$ is not isomorphic to the ring $[{\mathbb Z}(d^n)]$ (because of Eq.(\ref{17})).
The heart of the formalism is the local and global Fourier transforms.
We have shown that for some values of $d,n$ they are unitarily inequivalent to each other (proposition \ref{pro1}).

We have compared and contrasted the local phase space formalism with the global phase space formalism. Examples of this are:
\begin{itemize}
\item
Some of the local momentum states are the same as the global momentum states (proposition \ref{pro17}).
\item
Density matrices which have only diagonal elements with respect to the position basis, have the same local and global Wigner function (proposition \ref{pro22}).
The difference between local and global Wigner functions, is contained entirely in the off-diagonal elements.
\item
We have calculated the time evolution in terms of both  local variables and also global variables (section \ref{GGG})

\item
Classical and quantum correlations have been described in the local formalism
with the matrices $R_L$, $\widetilde R_L$ and in the global formalism with the matrices  $R_G$, $\widetilde R_G$. 

\end{itemize}

The formalism could be used in the general area of Fast Fourier transforms (in a quantum or even classical context).
For example, a link between the present formalism (in some special cases) and the Cooley-Tukey formalism has been discussed in section \ref{sec34}.

The work is a contribution to the various approaches for multipartite systems.
Unitary equivalence between the local and global Fourier transform (Eq.(\ref{59})), implies that the distinction between the concept of a multipartite system and that of a single system is weak. 
Unitary inequivalence (Eq.(\ref{60}) ) implies that the concept of a multipartite system is fundamentally different from that of  a single quantum system.

\subsection*{Conflict of interest and data availability statement}
We have no conflicts of interest to disclose.

No data were used in this paper, and therefore data availability is not applicable.

\newpage
 \begin{table}[h]
 \begin{math}
\begin{blockarray}{ccccccccccc}
    &  &  &  &  &  & \{$$\beta_0$$,$$\beta_1$$\} &  &  &  &  \\\cline{2-11}
\begin{block}{c|c|c|c|c|c|c|c|c|c|c|}
 & &\{-1,-1\} & \{0,-1\} & \{1,-1\} & \{-1,0\} & \{0,0\} & \{1,0\} & \{-1,1\} & \{0,1\} & \{1,1\} \\\cline{2-11}
      & \{-1,-1\} & 0 & 0 & 0 & 0 & 0 & 0 & 0 & 0 & 0 \\ 
         & \{0,-1\} & 0 & 0 & 0 & 0 & 0 & 0 & 0 & 0 & 0 \\ 
         & \{1,-1\} & 0 & 0 & 0 & 0 & 0 & 0 & 0 & 0 & 0 \\ 
         & \{-1,0\} & 0 & 0 & 0 & 0 & 0 & 0 & 0 & 0 & 0 \\ 
        \{$$\alpha_0$$,$$\alpha_1$$\} & \{0,0\} & 0 & 0 & 0 & 0 & 0 & 0 & 0 & 0 & 0 \\ 
         & \{1,0\} & 0 & 0 & 0 & 0 & 0 & 0 & 0 & 0 & 0 \\ 
         & \{-1,1\} & 0.4491 & -0.2931 & 0.4491 & 0.4491 & -0.2931 & 0.4491 & 0.4491 & -0.2931 & 0.4491 \\ 
         & \{0,1\} & -0.2931 & 0.844 & -0.2931 & -0.2931 & 0.844 & -0.2931 & -0.2931 & 0.844 & -0.2931 \\ 
         & \{1,1\} & 0.844 & 0.4491 & 0.844 & 0.844 & 0.4491 & 0.844 & 0.844 & 0.4491 & 0.844 \\  \cline{2-11}
\end{block}
\end{blockarray}
 \end{math}
 \caption{The local Wigner function ${W}_L(\{\alpha_0,\alpha_1;\beta_0,\beta_1\}|\rho_G)$ for the density matrix $\rho_G$ in Eq.(\ref{rholg})
 with $\widehat j=4$ and $d=3$, $n=2$.}
 \label{t1}
 \end{table}

 \begin{table}[h]
 \begin{adjustbox}{width=\columnwidth,center}
\begin{math}
\begin{blockarray}{ccccccccccc}
     &  &  &  &  &  & $$\widehat{\beta}=(\beta_0,\beta_1)$$ &  &  &  &  \\\cline{2-11}
\begin{block}{c|c|c|c|c|c|c|c|c|c|c|}
&  & \widehat{-4}=(-1,-1) & \widehat{-3}=(0,-1) & \widehat{-2}=(1,-1) & \widehat{-1}=(1,0)& \widehat{0} =(0,0)& \widehat{1}=(1,0) & \widehat{2}=(-1,1) & \widehat{3}=(0,1) & \widehat{4}=(1,1) \\\cline{2-11}
      & \widehat{-4}=(-1,-1) & 0 & 0 & 0 & 0 & 0 & 0 & 0 & 0 & 0 \\ 
         & \widehat{-3}=(0,-1) & 0 & 0 & 0 & 0 & 0 & 0 & 0 & 0 & 0 \\ 
         & \widehat{-2}=(1,-1) & -0.2931 & 0.844 & -0.2931 & -0.2931 & 0.844 & -0.2931 & -0.2931 & 0.844 & -0.2931 \\ 
         & \widehat{-1}=(-1,0) & 0 & 0 & 0 & 0 & 0 & 0 & 0 & 0 & 0 \\ 
        $$\widehat{\alpha}=(\alpha_0,\alpha_1)$$ & \widehat{0}=(0,0) & 0 & 0 & 0 & 0 & \widehat{0} & 0 & 0 & 0 & 0 \\ 
         & \widehat{1}=(1,0) & 0.4491 & -0.2931 & 0.4491 & 0.4491 & -0.2931 & 0.4491 & 0.4491 & -0.2931 & 0.4491 \\ 
         & \widehat{2}=(-1,1) & 0 & 0 & 0 & 0 & 0 & 0 & 0 & 0 & 0 \\ 
         & \widehat{3}=(0,1) & 0 & 0 & 0 & 0 & 0 & 0 & 0 & 0 & 0 \\ 
         & \widehat{4}=(1,1) & 0.844 & 0.4491 & 0.844 & 0.844 & 0.4491 & 0.844 & 0.844 & 0.4491 & 0.844 \\  \cline{2-11}
\end{block}
\end{blockarray}
\end{math}
\end{adjustbox}
 \caption{The global Wigner function ${W}_G(\widehat{\alpha}, \widehat{\beta}|\rho_L)$ for the density matrix $\rho_L$ in Eq.(\ref{rholg}), with $\widehat j=4$ and $d=3$, $n=2$.}
 \label{t2}
 \end{table}

\begin{table}[h]
 \begin{math}
\begin{blockarray}{ccccccccccc}
    &  &  &  &  &  & $$\{\delta_0$$,$$\delta_1\}$$ &  &  &  &  \\\cline{2-11}
\begin{block}{c|c|c|c|c|c|c|c|c|c|c|}
& &\{-1,-1\} & \{0,-1\} & \{1,-1\} & \{-1,0\} & \{0,0\} & \{1,0\} & \{-1,1\} & \{0,1\} & \{1,1\} \\\cline{2-11}
      & \{-1,-1\} & 0 & 0 & -0.1227 & 0.128 & 0 & 0 & 0 & 0.0106 & 0 \\ 
         & \{0,-1\} & 0 & 0 & -0.1227 & 0.128 & 0 & 0 & 0 & 0.0106 & 0 \\ 
         & \{1,-1\} & 0 & 0 & 0.9954 & 0.994 & 0 & 0 & 0 & 0.9788 & 0 \\ 
         & \{-1,0\} & 0 & 0 & -0.1227 & 0.128 & 0 & 0 & 0 & 0.0106 & 0 \\ 
        $$\{\gamma_0$$,$$\gamma_1\}$$ & \{0,0\} & 0 & 0 & 0.9954 & 0.994 & 0 & 0 & 0 & 0.9788 & 0 \\ 
         & \{1,0\} & 0 & 0 & -0.1227 & 0.128 & 0 & 0 & 0 & 0.0106 & 0 \\ 
         & \{-1,1\} & 0 & 0 & 0.9954 & 0.994 & 0 & 0 & 0 & 0.9788 & 0 \\ 
         & \{0,1\} & 0 & 0 & -0.1227 & 0.128 & 0 & 0 & 0 & 0.0106 & 0 \\ 
         & \{1,1\} & 0 & 0 & -0.1227 & 0.128 & 0 & 0 & 0 & 0.0106 & 0 \\ \cline{2-11}
\end{block}
\end{blockarray}
\end{math}
 \caption{The local Wigner function $W_L(\{\gamma_0,\gamma_1;\delta_0, \delta_1\}|\rho)$ for the density matrix $\rho$ in Eq.(\ref{68}).}
 \label{t3}
 \end{table}

 \begin{table}[h]
 \begin{math}
\begin{blockarray}{ccccccccccc}
    &  &  &  &  &  & $$\{\beta_0$$,$$\beta_1\}$$ &  &  &  &  \\\cline{2-11}
\begin{block}{c|c|c|c|c|c|c|c|c|c|c|}
 & &\{-1,-1\} & \{0,-1\} & \{1,-1\} & \{-1,0\} & \{0,0\} & \{1,0\} & \{-1,1\} & \{0,1\} & \{1,1\} \\\cline{2-11}
     & \{-1,-1\} & 0.067-0.0295i & 0 & 0 & 0 & -0.125+0.0722i & 0 & 0 & 0 & 0.067-0.0295i \\ 
         & \{0,-1\} & -0.059+0.0432i & 0 & 0 & 0 & 0.125-0.0722i & 0 & 0 & 0 & -0.059+0.0432i \\ 
         & \{1,-1\} & -0.4921-0.8523i & 0 & 0 & 0 & -0.5-0.866i & 0 & 0 & 0 & -0.4921-0.8523i \\ 
         & \{-1,0\} & -0.0079-0.0727i & 0 & 0 & 0 & 0.1443i & 0 & 0 & 0 & -0.0079-0.0727i \\ 
        $$\{\alpha_0$$,$$\alpha_1\}$$ & \{0,0\} & 0.9841 & 0 & 0 & 0 & 1 & 0 & 0 & 0 & 0.9841 \\ 
         & \{1,0\} & -0.0079+0.0727i & 0 & 0 & 0 & -0.1443i & 0 & 0 & 0 & -0.0079+0.0727i \\ 
         & \{-1,1\} & -0.4921+0.8523i & 0 & 0 & 0 & -0.5+0.866i & 0 & 0 & 0 & -0.4921+0.8523i \\ 
         & \{0,1\} & -0.059-0.0432i & 0 & 0 & 0 & 0.125+0.0722i & 0 & 0 & 0 & -0.059-0.0432i \\ 
         & \{1,1\} & 0.067+0.0295i & 0 & 0 & 0 & -0.125-0.0722i & 0 & 0 & 0 & 0.067+0.0295i \\  \cline{2-11}
\end{block}
\end{blockarray}
\end{math}
 \caption{The local Weyl function $\widetilde{W}_L(\{\alpha _0,\alpha_1;\beta _0,\beta_1\}|\rho)$ for the density matrix $\rho$ in Eq.(\ref{68})} 
 \label{t4}
 \end{table}

 \begin{table}[h]
 \begin{math}
\begin{blockarray}{ccccccccccc}
    &  &  &  &  &  & \{$$\delta_0$$,$$\delta_1$$\} &  &  &  &  \\\cline{2-11}
\begin{block}{c|c|c|c|c|c|c|c|c|c|c|}
 & &\{-1,-1\} & \{0,-1\} & \{1,-1\} & \{-1,0\} & \{0,0\} & \{1,0\} & \{-1,1\} & \{0,1\} & \{1,1\} \\\cline{2-11}
        & \{-1,-1\} & -0.1042 & -0.0833 & -0.1852 & -0.0456 & -0.1389 & -0.1042 & -0.1389 & -0.1005 & -0.0833 \\ 
         & \{0,-1\} & -0.1042 & -0.0833 & -0.1852 & -0.0456 & -0.1389 & -0.1042 & -0.1389 & -0.1005 & -0.0833 \\ 
         & \{1,-1\} & -0.1042 & -0.0833 & 0.9329 & 0.8204 & -0.1389 & -0.1042 & -0.1389 & 0.8677 & -0.0833 \\ 
         & \{-1,0\} & -0.1042 & -0.0833 & -0.1852 & -0.0456 & -0.1389 & -0.1042 & -0.1389 & -0.1005 & -0.0833 \\ 
        \{$$\gamma_0$$,$$\gamma_1$$\} & \{0,0\} & -0.1042 & -0.0833 & 0.9329 & 0.8204 & -0.1389 & -0.1042 & -0.1389 & 0.8677 & -0.0833 \\ 
         & \{1,0\} & -0.1042 & -0.0833 & -0.1852 & -0.0456 & -0.1389 & -0.1042 & -0.1389 & -0.1005 & -0.0833 \\ 
         & \{-1,1\} & -0.1042 & -0.0833 & 0.9329 & 0.8204 & -0.1389 & -0.1042 & -0.1389 & 0.8677 & -0.0833 \\ 
         & \{0,1\} & -0.1042 & -0.0833 & -0.1852 & -0.0456 & -0.1389 & -0.1042 & -0.1389 & -0.1005 & -0.0833 \\ 
         & \{1,1\} & -0.1042 & -0.0833 & -0.1852 & -0.0456 & -0.1389 & -0.1042 & -0.1389 & -0.1005 & -0.0833 \\  \cline{2-11}
\end{block}
\end{blockarray}
\end{math}
 
 \caption{The matrix ${R_L}(\{\gamma_0,\gamma_1;\delta_0,\delta _1\}|\rho)$ for the density matrix $\rho$ in Eq.(\ref{68}).}
 \label{t5}
 \end{table}

 \begin{table}[h]
 \begin{math}
\begin{blockarray}{ccccccccccc}
    &  &  &  &  &  & $$\{\beta_0$$,$$\beta_1\}$$ &  &  &  &  \\\cline{2-11}
\begin{block}{c|c|c|c|c|c|c|c|c|c|c|}
& &\{-1,-1\} & \{0,-1\} & \{1,-1\} & \{-1,0\} & \{0,0\} & \{1,0\} & \{-1,1\} & \{0,1\} & \{1,1\} \\\cline{2-11}
     & \{-1,-1\} & 0.067-0.0295i & 0 & 0 & 0 & -0.1354+0.0541i & 0 & 0 & 0 & 0.067-0.0295i \\ 
         & \{0,-1\} & -0.059+0.0432i & 0 & 0 & 0 & 0 & 0 & 0 & 0 & -0.059+0.0432i \\ 
         & \{1,-1\} & -0.4921-0.8523i & 0 & 0 & 0 & -0.4896-0.848i & 0 & 0 & 0 & -0.4921-0.8523i \\ 
         & \{-1,0\} & -0.0079-0.0727i & 0 & 0 & 0 & 0 & 0 & 0 & 0 & -0.0079-0.0727i \\ 
        $$\{\alpha_0$$,$$\alpha_1\}$$ & \{0,0\} & 0.9841 & 0 & 0 & 0 & 0 & 0 & 0 & 0 & 0.9841 \\ 
         & \{1,0\} & -0.0079+0.0727i & 0 & 0 & 0 & 0 & 0 & 0 & 0 & -0.0079+0.0727i \\ 
         & \{-1,1\} & -0.4921+0.8523i & 0 & 0 & 0 & -0.4896+0.848i & 0 & 0 & 0 & -0.4921+0.8523i \\ 
         & \{0,1\} & -0.059-0.0432i & 0 & 0 & 0 & 0 & 0 & 0 & 0 & -0.059-0.0432i \\ 
         & \{1,1\} & 0.067+0.0295i & 0 & 0 & 0 & -0.1354-0.0541i & 0 & 0 & 0 & 0.067+0.0295i \\  \cline{2-11}
\end{block}
\end{blockarray}
 \end{math}
 \caption{The matrix $\widetilde{R_L}(\{\alpha_0,\alpha_1;\beta_0,\beta _1\}|\rho)$ for the density matrix $\rho$ in Eq.(\ref{68}).}
 \label{t6}
 \end{table}

 \begin{table}[h]
  \begin{adjustbox}{width=\columnwidth,center}
 \begin{math}
\begin{blockarray}{ccccccccccc}
    &  &  &  &  &  & $$\widehat{\delta} =(\delta_0,\delta_1)$$ &  &  &  &  \\\cline{2-11}
\begin{block}{c|c|c|c|c|c|c|c|c|c|c|}
        &  & \widehat{-4}=(-1,-1) & \widehat{-3}=(0,-1) & \widehat{-2}=(1,-1) & \widehat{-1}=(1,0)& \widehat{0} =(0,0)& \widehat{1}=(1,0) & \widehat{2}=(-1,1) & \widehat{3}=(0,1) & \widehat{4}=(1,1) \\\cline{2-11}
     & \widehat{-4}=(-1,-1) & 0.1003 & 0 & 0.25 & 0.4167 & 0 & 0.1294 & 0 & -0.2732 & 0 \\ 
         & \widehat{-3}=(0,-1) & -0.2887 & 0 & 0.25 & 0.4167 & 0 & -0.3727 & 0 & 0.0106 & 0 \\ 
         & \widehat{-2}=(1,-1) & 0.4423 & 0 & 0.25 & 0.4167 & 0 & 0.571 & 0 & 0.4454 & 0 \\ 
         & \widehat{-1}=(-1,0) & -0.5425 & 0 & 0.25 & 0.4167 & 0 & -0.7004 & 0 & 0.8278 & 0 \\ 
       $$\widehat{\gamma} =(\gamma_0,\gamma_1)$$ & \widehat{0}=(0,0) & 0.5774 & 0 & 0.25 & 0.4167 & 0 & 0.7454 & 0 & 0.9788 & 0 \\ 
         & \widehat{1}=(1,0) & -0.5425 & 0 & 0.25 & 0.4167 & 0 & -0.7004 & 0 & 0.8278 & 0 \\ 
         & \widehat{2}=(-1,1) & 0.4423 & 0 & 0.25 & 0.4167 & 0 & 0.571 & 0 & 0.4454 & 0 \\ 
         & \widehat{3}=(0,1) & -0.2887 & 0 & 0.25 & 0.4167 & 0 & -0.3727 & 0 & 0.0106 & 0 \\ 
         & \widehat{4}=(1,1) & 0.1003 & 0 & 0.25 & 0.4167 & 0 & 0.1294 & 0 & -0.2732 & 0 \\  \cline{2-11}
\end{block}
\end{blockarray}
 \end{math}
  \end{adjustbox}
 \caption{The global Wigner function $W_G(\widehat{\gamma}, \widehat{\delta}|\rho)$ for the density matrix $\rho$ in Eq.(\ref{68}).}
 \label{t7}
 \end{table}
 
  \begin{table}[h]
 \begin{adjustbox}{width=\columnwidth,center}
 \begin{math}
\begin{blockarray}{ccccccccccc}
    &  &  &  &  &  & $$\widehat{\beta}=(\beta_0,\beta_1)$$ &  &  &  &  \\\cline{2-11}
\begin{block}{c|c|c|c|c|c|c|c|c|c|c|}
&  & \widehat{-4}=(-1,-1) & \widehat{-3}=(0,-1) & \widehat{-2}=(1,-1) & \widehat{-1}=(1,0)& \widehat{0} =(0,0)& \widehat{1}=(1,0) & \widehat{2}=(-1,1) & \widehat{3}=(0,1) & \widehat{4}=(1,1) \\\cline{2-11}
     & \widehat{-4}=(-1,-1) & -0.3001-0.4118i & 0 & 0 & -0.1614-0.2795i & -0.3667-0.3069i & -0.1614-0.2795i & 0 & 0 & -0.3001-0.4118i \\ 
         & \widehat{-3}=(0,-1) & -0.3307-0.0727i & 0 & 0 & 0.3227 & 0.1443i & 0.3227 & 0 & 0 & -0.3307-0.0727i \\
         & \widehat{-2}=(1,-1) & 0.2859-0.5526i & 0 & 0 & -0.1614+0.2795i & -0.3292+0.7845i & -0.1614+0.2795i & 0 & 0 & 0.2859-0.5526i \\ 
         & \widehat{-1}=(-1,0) & 0.0142-0.1408i & 0 & 0 & -0.1614-0.2795i & 0.1959+0.2254i & -0.1614-0.2795i & 0 & 0 & 0.0142-0.1408i \\ 
        $$\widehat{\alpha}=(\alpha_0,\alpha_1)$$ & \widehat{0}=(0,0) & 0.6614 & 0 & 0 & 0.3227 & 1 & 0.3227 & 0 & 0 & 0.6614 \\ 
         & \widehat{1}=(1,0) & 0.0142+0.1408i & 0 & 0 & -0.1614+0.2795i & 0.1959-0.2254i & -0.1614+0.2795i & 0 & 0 & 0.0142+0.1408i \\ 
         & \widehat{2}=(-1,1) & 0.2859+0.5526i & 0 & 0 & -0.1614-0.2795i & -0.3292-0.7845i & -0.1614-0.2795i & 0 & 0 & 0.2859+0.5526i \\ 
         & \widehat{3}=(0,1) & -0.3307+0.0727i & 0 & 0 & 0.3227 & -0.1443i & 0.3227 & 0 & 0 & -0.3307+0.0727i \\ 
         & \widehat{4}=(1,1) & -0.3001+0.4118i & 0 & 0 & -0.1614+0.2795i & -0.3667+0.3069i & -0.1614+0.2795i & 0 & 0 & -0.3001+0.4118i \\  \cline{2-11}
\end{block}
\end{blockarray}
 \end{math}
 \end{adjustbox}
 \caption{The global Weyl function $\widetilde{W}_G(\widehat{\alpha}, \widehat{\beta}|\rho)$ for the density matrix $\rho$ in Eq.(\ref{68}).}
 \label{t8}
 \end{table}
 
   \begin{table}[h]
 \begin{adjustbox}{width=\columnwidth,center}
 \begin{math}
\begin{blockarray}{ccccccccccc}
    &  &  &  &  &  & $$\widehat{\delta} =(\delta_0,\delta_1)$$ &  &  &  &  \\\cline{2-11}
\begin{block}{c|c|c|c|c|c|c|c|c|c|c|}
&  & \widehat{-4}=(-1,-1) & \widehat{-3}=(0,-1) & \widehat{-2}=(1,-1) & \widehat{-1}=(1,0)& \widehat{0} =(0,0)& \widehat{1}=(1,0) & \widehat{2}=(-1,1) & \widehat{3}=(0,1) & \widehat{4}=(1,1) \\\cline{2-11}
       & \widehat{-4}=(-1,-1) & -0.0039 & -0.0833 & 0.1875 & 0.2431 & -0.1389 & 0.0253 & -0.1389 & -0.3843 & -0.0833 \\ 
         & \widehat{-3}=(0,-1) & -0.3928 & -0.0833 & 0.1875 & 0.2431 & -0.1389 & -0.4768 & -0.1389 & -0.1005 & -0.0833 \\ 
         & \widehat{-2}=(1,-1) & 0.3381 & -0.0833 & 0.1875 & 0.2431 & -0.1389 & 0.4668 & -0.1389 & 0.3343 & -0.0833 \\ 
         & \widehat{-1}=(-1,0) & -0.6467 & -0.0833 & 0.1875 & 0.2431 & -0.1389 & -0.8046 & -0.1389 & 0.7167 & -0.0833 \\ 
        $$\widehat{\gamma} =(\gamma_0,\gamma_1)$$ & \widehat{0}=(0,0) & 0.4732 & -0.0833 & 0.1875 & 0.2431 & -0.1389 & 0.6412 & -0.1389 & 0.8677 & -0.0833 \\ 
         & \widehat{1}=(1,0) & -0.6467 & -0.0833 & 0.1875 & 0.2431 & -0.1389 & -0.8046 & -0.1389 & 0.7167 & -0.0833 \\ 
         & \widehat{2}=(-1,1) & 0.3381 & -0.0833 & 0.1875 & 0.2431 & -0.1389 & 0.4668 & -0.1389 & 0.3343 & -0.0833 \\ 
         & \widehat{3}=(0,1) & -0.3928 & -0.0833 & 0.1875 & 0.2431 & -0.1389 & -0.4768 & -0.1389 & -0.1005 & -0.0833 \\ 
         & \widehat{4}=(1,1) & -0.0039 & -0.0833 & 0.1875 & 0.2431 & -0.1389 & 0.0253 & -0.1389 & -0.3843 & -0.0833 \\  \cline{2-11}
\end{block}
\end{blockarray}
 \end{math}
 \end{adjustbox}
 \caption{The matrix $R_G(\widehat{\gamma}, \widehat{\delta}|\rho)$ for the density matrix $\rho$ in Eq.(\ref{68}).}
 \label{t9}
 \end{table}
 
 \begin{table}[h]
 \begin{adjustbox}{width=\columnwidth,center}
 \begin{math}
\begin{blockarray}{ccccccccccc}
    &  &  &  &  &  & $$\widehat{\beta}=(\beta_0,\beta_1)$$ &  &  &  &  \\\cline{2-11}
\begin{block}{c|c|c|c|c|c|c|c|c|c|c|}
&  & \widehat{-4}=(-1,-1) & \widehat{-3}=(0,-1) & \widehat{-2}=(1,-1) & \widehat{-1}=(1,0)& \widehat{0} =(0,0)& \widehat{1}=(1,0) & \widehat{2}=(-1,1) & \widehat{3}=(0,1) & \widehat{4}=(1,1) \\\cline{2-11}
       & \widehat{-4}=(-1,-1) & -0.3001-0.4118i & 0 & 0 & -0.1614-0.2795i & -0.3342-0.3351i & -0.1614-0.2795i & 0 & 0 & -0.3001-0.4118i \\ 
         & \widehat{-3}=(0,-1) & -0.3307-0.0727i & 0 & 0 & 0.3227 & 0 & 0.3227 & 0 & 0 & -0.3307-0.0727i \\ 
         & \widehat{-2}=(1,-1) & 0.2859-0.5526i & 0 & 0 & -0.1614+0.2795i & -0.3735+0.7316i & -0.1614+0.2795i & 0 & 0 & 0.2859-0.5526i \\ 
         & \widehat{-1}=(-1,0) & 0.0142-0.1408i & 0 & 0 & -0.1614-0.2795i & 0.0827+0.2729i & -0.1614-0.2795i & 0 & 0 & 0.0142-0.1408i \\ 
        $$\widehat{\alpha}=(\alpha_0,\alpha_1)$$ & \widehat{0}=(0,0) & 0.6614 & 0 & 0 & 0.3227 & 0 & 0.3227 & 0 & 0 & 0.6614 \\ 
         & \widehat{1}=(1,0) & 0.0142+0.1408i & 0 & 0 & -0.1614+0.2795i & 0.0827-0.2729i & -0.1614+0.2795i & 0 & 0 & 0.0142+0.1408i \\ 
         & \widehat{2}=(-1,1) & 0.2859+0.5526i & 0 & 0 & -0.1614-0.2795i & -0.3735-0.7316i & -0.1614-0.2795i & 0 & 0 & 0.2859+0.5526i \\ 
         & \widehat{3}=(0,1) & -0.3307+0.0727i & 0 & 0 & 0.3227 & 0 & 0.3227 & 0 & 0 & -0.3307+0.0727i \\ 
         & \widehat{4}=(1,1) & -0.3001+0.4118i & 0 & 0 & -0.1614+0.2795i & -0.3342+0.3351i & -0.1614+0.2795i & 0 & 0 & -0.3001+0.4118i \\  \cline{2-11}
\end{block}
\end{blockarray}
 \end{math}
 \end{adjustbox}
 \caption{The matrix $\widetilde{R_G}(\widehat{\alpha}, \widehat{\beta}|\rho)$ for the density matrix $\rho$ in Eq.(\ref{68}).}
 \label{t10}
 \end{table}


\begin{thebibliography}{999}

\bibitem{H}
R. Horodecki, P. Horodecki, M. Horodecki, K. Horodecki, `Quantum entanglement', Rev. Mod. Phys. 81, 865 (2009)
\bibitem{V1}
A. Vourdas, `Finite and profinite quantum systems' (Springer, Berlin, 2017)
\bibitem{V11}
T. Durt, B.G. Englert, I. Bengtsson, K. Zyczkowski, `On mutually unbiased bases', Int. J. Quantum Comput. 8, 535 (2010)
\bibitem{V2}
A. Vourdas, `Multipartite quantum systems: an approach based on Markov matrices and the Gini index', J. Phys A54, 185201 (2021)
\bibitem{B1}
J.H. McClellan, C.M. Rader, `Number theory in digital signal processing' (Prentice Hall, New Jersey, 1979)
\bibitem{B2}
R.E. Blahut `Fast algorithms for digital signal processing' Addison-Wesley, Reading Mass, 1985)
\bibitem{B3}
D.F. Elliott, K.R. Rao, `Fast transforms' (Academic Press, London, 1982)
\bibitem{EV0}
M. Horibe, A. Takami, T. Hashimoto, A. Hayashi, `Existence of the Wigner function with correct marginal distributions along tilted lines on a lattice', Phys. Rev. A65, 032105 (2002)
\bibitem{EV1}
T. Durt, `About mutually unbiased bases in even and odd prime power dimensions', J. Phys. A38, 5267 (2005)
\bibitem{EV2}
J. Zak, `Doubling feature of the Wigner function: finite phase space', J. Phys. A44, 345305 (2011)
\bibitem{T}
A. Terras `Fourier analysis on finite groups and applications (Cambridge Univ. Press, Cambridge, 1999)
\bibitem{G}
I.J. Good, `The relationship between two fast Fourier transforms', IEEE Transactions on computers C-20, 310 (1971)
\bibitem{V3}
A. Vourdas, `Factorisation in finite quantum systems', J. Phys. A36, 5645 (2003)
\bibitem{E1}
T.G. Gerasimova, `Unitary similarity to a normal matrix', Linear Algebra Appl. 436, 3777 (2012)
\bibitem{E2}
H. Shapiro, `A survey of canonical forms and invariants for unitary similarity', Linear Algebra Appl. 147, 101 (1991)



\end{thebibliography}
\end{document}